\documentclass[12pt]{iopart}

\usepackage[pdftex]{graphicx}
\usepackage{amsthm}
\usepackage{amssymb}
\usepackage{amsfonts}
\usepackage{braket}
\usepackage[mathcal]{eucal}
\usepackage{bm}

\newcommand{\eqref}[1]{(\ref{#1})}
\newcommand{\Ref}{\eqref}

\renewcommand{\le}{\leqslant}
\renewcommand{\ge}{\geqslant}
\renewcommand{\AA}{\mathcal{A}}
\newcommand{\CC}{\mathcal{C}}
\newcommand{\HH}{\mathcal{H}}
\newcommand{\MM}{\mathcal{M}}
\newcommand{\RR}{\mathcal{R}}
\newcommand{\TT}{\mathcal{T}}
\newcommand{\UU}{\mathcal{U}}

\renewcommand{\a}{\alpha}

\renewcommand{\d}{\delta}

\newcommand{\ep}{\varepsilon}

\newcommand{\la}{\lambda}
\newcommand{\lam}{\lambda}

\newcommand{\s}{\sigma}
\newcommand{\ta}{\tau}

\newcommand{\su}{\mathfrak{su}}
\newcommand{\SU}{\mathrm{SU}}

\newcommand{\paren}[1]{{\left( #1 \right)}}
\newcommand{\brac}[1]{{\left\{ #1 \right\}}}
\newcommand{\normalbrac}[1]{{\{ #1 \}}}
\newcommand{\brak}[1]{{\left[ #1 \right]}}
\newcommand{\angl}[1]{\left\langle#1\right\rangle}
\newcommand{\f}{\frac}
\newcommand{\q}{\quad}

\newcommand{\nn}{\nonumber \\ }

\newcommand{\diff}[2][]{\f{\mathrm{d}#1}{\mathrm{d}#2}}
\newcommand{\pdiff}[2][]{\f{\partial #1}{\partial #2}}
\newcommand{\MP}{MP}
\newcommand{\QMP}{MP-QB}
\newcommand{\hpmp}{H_{\mathrm{{\MP}}}}
\newcommand{\bmat}{\left(\begin{array}}
    \newcommand{\emat}{\end{array}\right)}
\newcommand{\matx}[1]{\small\bmat{ccc}#1\emat{}}

\newcommand{\Span}{\mathrm{span}\,}
\newcommand{\binom}[2]{\small
      \bmat{c}
      \hspace{-0.5em} #1 \hspace{-0.5em}\\
      \hspace{-0.5em} #2 \hspace{-0.5em}
      \emat{}}

\begin{document}
\title{A general formulation of time-optimal quantum control and optimality of singular protocols}
\author{Hiroaki Wakamura}
\ead{hwakamura@rk.phys.keio.ac.jp}
\address{Department of Physics, Keio University, Yokohama 223-8522, Japan}
\author{Tatsuhiko Koike}
\ead{koike@phys.keio.ac.jp}
\address{Department of Physics, Keio University, Yokohama 223-8522, Japan}
\address{Quantum Computing Center,
  Keio University, Yokohama 223-8522, Japan}
\address{Research and Education Center for Natural Sciences,
  Keio University, Yokohama 223-8521, Japan}
\date{\today}

\begin{abstract}
  We present a general theoretical framework for finding the
  time-optimal unitary evolution of the quantum systems when the
  Hamiltonian is subject to arbitrary constraints. Quantum
  brachistochrone (QB) is such a framework
  based on the variational principle,
  whose drawback is that it deals with equality
  constraints only.
  While inequality constraints can be reduced to
  equality ones in some situations,
  there are situations where they
  cannot, especially when a drift field is present in the
  Hamiltonian. The drift which we cannot
  control
  appears in a wide range of systems.
  We first develop a framework based on Pontryagin's maximum
  principle ({\MP}) 
  in order to deal with inequality constraints as well. 
  The new framework contains QB as a special case,
  and their detailed correspondence is given.
  Second,
  using this framework,
  we discuss general relations among
  the drift, the singular controls, and
  the inequality constraints.
  The singular controls
  are those that
  satisfy {\MP} trivially
  so as to cause a trouble in determining the optimal protocol.
  Third,
  to overcome this issue,
  we derive an additional necessary condition
  for a singular protocol to be optimal
  by
  applying the generalized Legendre-Clebsch condition.
  This condition in particular reveals
  the physical meaning of singular controls.
  Finally,
  we demonstrate how our framework and results work in some examples.
\end{abstract}

\maketitle

\section{Introduction}
\label{sec:intro}

Control of quantum systems
is important
both in applications and
in fundamental physics.
In applications,
development of quantum control
will help
perform a variety of tasks
in quantum technologies
such as quantum computation
and quantum cryptography.
In the fundamental aspect,
control theory can clarify the limitations
on our abilities to complete a certain task
in quantum mechanics
and provides a way of understanding quantum mechanics
from an engineering point of view.
An important subject in this line
is the time-optimal
quantum control~\cite{Kha.Bro.Gla2001a,Kha.Gla2001},
which realizes the desired unitary
evolution in the least possible time.
Time-optimal quantum control is useful
in a variety of applications,
in that
it saves time cost
and enables manipulations
of quantum systems in
short decoherence time.
In fact, some studies
aim to consider
time-optimal quantum control
for practical purposes,
including efficient
design of elementary gates and
subroutines in quantum
computation~\cite{Sch.Spo.Kha.Gla2005,Hua.Goa2014},
quantum error-correcting codes~\cite{Bur.Los.DiV.Smo1999},
and cooling of quantum systems~\cite{Wan.Vin.Str.Jac2011,Mac.Cer.Asp.Wie.Ple.Ret2012}.
On the other hand, time-optimal
quantum control gives the physically
attainable minimal time
in the fundamental sense
to realize prescribed quantum evolutions.
Such minimal time may also be an important
tool to estimate the complexity
of quantum
computation~\cite{Nie2006,Nie.Dow.Gu.Doh2006,Koi.Oku2010}.

There are many studies related
to time-optimal or time-efficient quantum control.
Khaneja \etal{}~\cite{Kha.Bro.Gla2001a,Kha.Gla2001}
and Zhang \etal{}~\cite{Zha.Val.Sas.Wha2003}
discussed time-optimal control
by using the Lie algebraic
approach.
They treated such situations
that some unitary operations
can be done instantaneously,
which is a good approximation
if the corresponding
interactions are sufficiently large.
\textit{Quantum brachistochrone} (QB)
is a general theory for
time-optimal quantum control
problems~\cite{Car.Hos.Koi.Oku2006,
  Car.Hos.Koi.Oku2007,Car.Hos.Koi.Oku2008}.
QB treats the situations where
the system Hamiltonian
can be changed under some
equality constraints.
Russell and Stepney~\cite{Rus.Ste2014,Rus.Ste2015},
and Brody and Meier~\cite{Bro.Mei2015a}
discuss  geometric methods for
deriving the time-optimal Hamiltonian
in the presence of a \textit{drift field},
which is a fixed, uncontrollable
interaction.
There are also many 
studies~\cite{Kir.Hoc.Spi2008,Boo2012,
  Bil2013,Gar.Gla.Sug2013,
  Rom2014,Alb.DA2015}
which focus
on unitary operations on
a qubit.
Among the above-mentioned researches, 
QB may be able to
serve as a standard theory
for time-optimal control
in quantum systems because of its generality and wide applicability,
and may provide a unified understanding of the results from other
studies.
For example, we will see later that
the results by Brody and Meier~\cite{Bro.Mei2015a}
can be shown in a simple manner by QB.
One can 
obtain analytical optimal 
solutions of QB for low dimensional
systems~\cite{Car.Hos.Koi.Oku2006,
  Car.Hos.Koi.Oku2007}.
Recently, experimental 
realizations of time-optimal
quantum control 
based on QB 
started to appear~\cite{Gen.Wu.Wan.Xu.Shi.Xie.Ron.Du2016b}.

However, there is a weakness in
the present QB. It cannot
treat problems with
inequality constraints on the
system Hamiltonian,
which are natural 
in usual physical systems.
The most important
inequality constraint is
the one that the system has
a finite energy bandwidth,
which is usually imposed
by the bounded capability of
the experimental
apparatus.
A previous study~\cite{Wan.All.Jac.Llo.Lup.Moh2015} shows
that this inequality constraint
is reduced to an equality
one in 
certain situations.
However,
it is not always the case
in a more general situation
where a drift field is present.
In fact,
Ref.~\cite{Heg2013} shows an example
of the time-optimal control problem
in which
the inequality constraint cannot be
reduced to the equality one.
Since we cannot apply
the present QB
to such problems,
we need to extend QB theory.
We also want to clarify in which situation
inequality constraints reduce to equality ones.

In this paper,
we
present
a more general time-optimal
control theory in quantum systems,
which can be applied to the problem
with inequality constraints.
Instead of directly extending
the present QB formulation,
we employ an alternative
approach
to get compatible
results with
QB.
Our theory is based on 
Pontryagin's maximum principle ({\MP}),
which is a non-traditional
variational calculus
for optimal control
under equality and inequality constraints.
In fact, {\MP} has been applied
in many
studies~\cite{Kir.Hoc.Spi2008,Boo2012,Gar.Gla.Sug2013,Rom2014,Alb.DA2015,Heg2013,Lap.Zha.Bra.Gla.Sug2010,Zha.Lap.Sug.Bra.Gla2011,Avi.Fis.Lon.Ger2014}
of time-optimal control
in quantum systems.
There, however, {\MP} is usually applied after individual quantum control
problems are
cast in a standard form in classical control theory
with real variables.
We will apply 
{\MP} to
time-optimal
quantum
control
in its general form 
so that we can discuss
the structure of time-optimal quantum
control itself
and that we can solve individual problems more simply and directly.

For the system with drift fields,
we have a special class of
solution for {\MP},
which satisfies
the necessary conditions provided by {\MP}
in a trivial way.
Such solutions are called
\textit{singular controls}~\cite{Alb.DA2015,
  Lap.Zha.Bra.Gla.Sug2010}.
Although we cannot determine
the optimal singular control uniquely
from {\MP},
we can
derive additional conditions
for the singular control to be optimal.
This can be done by applying the generalized
Legendre-Clebsch conditions~\cite{Rob1967}.
We will demonstrate
in some examples that 
the obtained conditions
determine, or at least restrict, the form
of singular optimal control.
The conditions are
useful not only in applications
but also in the interpretation
of the singular control.
We will see
that singular controls tend to
make use of the drift field.

The paper is organized as follows.
We first formulate the time-optimal
control problem in quantum systems in
Sec.~\ref{sec-setup}.
In Sec.~\ref{sec-qb}, we review the
theory of
QB
with some refinement
and then explain
some results
derived from QB.
We present a general framework
for the time-optimal control in
quantum systems by virtue of {\MP}
and clarify its relation to
QB in
Sec.~\ref{sec-pmp}.
In Sec.~\ref{sec-drift-ineq},
we show some conditions
for an inequality constraint
to be reduced into an equality one.
In Sec.~\ref{sec-pmp-singular},
we give the definition of
singular controls 
and provide some additional necessary conditions
for a singular control to be optimal.
In Sec.~\ref{sec-example},
we discuss optimality of
singular controls in some examples
by applying the developed theory and necessary conditions.
Section~\ref{sec-conclusion}
is devoted to conclusion and discussions.

\section{The problem}
\label{sec-setup}

We consider the problem of finding
the time-optimal control protocol
that generates a desired unitary
evolution
of a quantum system
in the least possible time.
There, we can design the time dependence of
the Hamiltonian
so as to realize the desired evolution
through the Schr\"odinger equation.
The optimal control protocol
largely depends on
the form of
available Hamiltonians, which is usually
restricted due to
the experimental or theoretical setting.
In this section, we shall
formulate the problem
and introduce some notations.

Consider a physical system
represented by an $N$-dimensional
Hilbert space $\HH$.
The group of unitary operators on
$\HH$ is identified with the $N$-dimensional
unitary group $\mathrm{U}(N)$ (via a choice of a basis on $\HH$).
We shall deal with 
the subgroup $\SU(N)$ instead of
$\mathrm{U}(N)$, neglecting the
global phase of unitary operator,
which does not affect the observables.
This is equivalent to think of
the Hamiltonian $H$ as an element of $\su(N)$,
i.e. a
traceless Hermitian operator.
Here, $\su(N)$ denotes the Lie algebra of
$\SU(N)$ with
the convention
$\SU(N)=\exp(-i \,\su(N))$.
We will often introduce an orthonormal basis
$\brac{\tau_j}_{j=1}^{N^2-1}$
of $\su(N)$
such that 
$\tr\brak{\tau_i\tau_j}=2\delta_{ij}$. 
Any Hermitian operator 
$A$ is then expanded as $A = \sum_j a^j\tau_j$ with real
coefficients $a^j$.

Let $\AA\subset\su(N)$ be the set of
\textit{available}
Hamiltonians
which are realizable under the given experimental or theoretical
setup.
We assume that
$\AA$ is time independent.
The set $\AA$ of available Hamiltonians
is usually defined by equality
and inequality constraints mathematically.

In these terms, the problem we want to solve is the following.
Let $(U_f,\AA)$ be a given pair,
where $U_f\in\SU(N)$ is the target unitary operator and
$\AA\subset\su(N)$ is the set of available Hamiltonians.
Find the least time $T\ge0$ and
the control $H(t)\in\AA$, $0\le t\le T$,
such that $U(0)=1$ and $U(T)=U_f$,
where the unitary operator $U(t)$ is driven by
the Schr\"odinger equation
\begin{eqnarray}
  i\diff{t}U(t) = H(t)U(t).
  \label{eq-sch}
\end{eqnarray}

We
give a simplest example
of the time-optimal control problem in
a qubit system.
We assume that a $z$-directional
magnetic field is fixed
and that we can manipulate a 
magnetic field in the $xy$ plane with 
the maximum magnitude $\Omega$.
The Hamiltonian is given by
\begin{eqnarray}
  H(t) = \omega_0\s^z + u^x(t)\s^x +u^y(t)\s^y,
  \quad
  (u^x)^2+(u^y)^2\le \Omega^2,
  \label{eq-simple-hamil-form}
\end{eqnarray}
where $\omega_0$ is a real constant.
Since this Hamiltonian can generate arbitrary
unitary operator in $\SU(2)$ with
suitable control variables $u^x(t)$ and $u^y(t)$,
we can consider the time-optimal control
for any $U_f \in \SU(2)$.
In this case,
we can write the set $\AA$ of
available Hamiltonians
as
\begin{eqnarray}
  \AA = \brac{\omega_0\s^z+H_c \;\middle | \;
    \tr\brak{H_c(t)\s^z}=0,\;
    \f12\tr \brak{H_c(t)^2}\le \Omega^2
  }.
\end{eqnarray}

\section{Quantum Brachistochrone}
\label{sec-qb}

Quantum brachistochrone
(QB)~\cite{Car.Hos.Koi.Oku2006,
  Car.Hos.Koi.Oku2007,Car.Hos.Koi.Oku2008}
is a theory of time-optimal control
on quantum systems.
In this section,
we review
QB theory but with a slightly different derivation,
which might be simpler.
The present derivation
is convenient in establishing a general
correspondence between QB and a theory based on {\MP}
given in the next section,
where the latter can treat inequality constraints.
We also discuss (in Sec.~\ref{qb-direct-cons})
some general results
which are directly obtained from QB,
including the solution of the quantum Zermelo
navigation~\cite{Rus.Ste2014, Bro.Mei2015a,Rus.Ste2015}.

\subsection{Formulation}

We assume that
the set $\AA$ of available Hamiltonians is
given by
\begin{eqnarray}
  \label{eq-qb-eqconstr}
  \AA=\brac{H\in\su(N)\;\middle|\; f^j(H)=0, \, j=1,\dots,p},
\end{eqnarray}
where
$f^j$ are real functions.
Then
we can formulate 
the time-optimality problem in the previous section 
as follows:
Minimize the functional $T= \int_0^T dt \;1$
of the Hamiltonian $H(t)$
and the unitary operator $U(t)$
under
constraints $f^j(H)=0$ and
the Schr\"odinger equation~\eqref{eq-sch}.
We employ the method of
Lagrange multipliers
and define an
action,
\begin{eqnarray}
  S := \int_0^T dt\;\paren{1+L_S+L_C},
  \label{eq-qb-action}
\end{eqnarray}
where
\begin{eqnarray}
  L_S
  \label{eq-qb-LS}
  &:= \tr\brak{F(t)\paren{i\dot{U}U^\dag-H(t)}}, \\
  L_C &:= \sum_j\lam_j(t) f^j(H),
  \label{eq-qb-LC}
\end{eqnarray}
and $F(t)\in \su(N)$ and $\lam_j(t)\in \mathbb{R}\;(j=0,1,\dots,p)$
are Lagrange multipliers.
In the present formulation,
the main part of the integrand in the action \eqref{eq-qb-action}
is simply 1 and
the variation of the action $S$
is explicitly taken
with respect to the final time $T$.
In the previous derivation~\cite{Car.Hos.Koi.Oku2006,
  Car.Hos.Koi.Oku2007,Car.Hos.Koi.Oku2008},
the variation of the total time is expressed
in an indirect manner
which is similar to that used
in variation of
arc length in differential geometry or in general relativity.%
\footnote{%
There the main part of the action was expressed in a
reparametrization-invariant manner
and
the variations were taken with respect to $U(t)$, $H(t)$, etc.
The mathematical treatment was similar to the derivation
of geodesics
from variation of arc length.
}
The term $L_S$ represents the constraints
that $U(t)$ must be related to
$H(t)$ by
the Schr\"odinger equation.
The term
$L_C$ represents the equality
constraints $f^j(H)=0$ imposed.

By the method of Lagrange multipliers,
we need $\delta S = 0$
under the variations
with respect to $H(t),U(t),F(t),\lam_j(t)$,
and $T$.
The condition that the final unitary operator $U(T)$
is not changed
by the variations imposes a condition for $\delta T$.
Namely, we have both
$U(T)=U_f$ before taking the variations
and $U(T+\delta T)+\delta U(T+\delta T)=U_f$
after taking the
variations.
Therefore, we
require
a condition at $t=T$,
\begin{eqnarray}
  \delta U(T) + \dot{U}(T)\delta T =0.
  \label{eq-terminal-condition}
\end{eqnarray}

\subsection{Result}
The result from the variational principle above is as follows, 
while 
its derivation is given in
the next subsection. 
If $H(t)$ is a time-optimal control,
then there exist $\lam_j(t)\in\mathbb{R}$
for $j=0,1,\dots,p$
such that
the operator\footnote{
  The partial differentiation is
  defined so that
  $\tr\brak{ (\partial f^j(H)/\partial H) A}
  =
  \lim_{\ep\to0} \paren{f^j(H+\ep A)-f^j(H)}/\ep$.
}
\begin{eqnarray}
  F(t) =
  \sum_j \lam_j(t)
  \f{\partial f^j(H)}{\partial H},
  \label{eq-QB-F}
\end{eqnarray}
satisfies the \textit{quantum brachistochrone equation}
\begin{eqnarray}
  i\dot{F}(t) = \brak{H(t),F(t)},
  \label{eq-QBeq}
\end{eqnarray}
and
\begin{eqnarray}
  \tr\brak{H(t)F(t)} = 1.
  \label{eq-QB-surface}
\end{eqnarray}
Therefore, 
we can obtain the time-optimal protocol $H(t)$ for $(U_f,\AA)$ 
by solving 
the equations 
\eqref{eq-QB-F}, 
\eqref{eq-QBeq}, 
and 
\eqref{eq-QB-surface} 
with the boundary conditions $U(0)=1$ and $U(T)=U_f$. 

The statement above is the same as that of 
Ref.~\cite{Car.Hos.Koi.Oku2007}, except that 
eq.~\eqref{eq-QB-surface} 
is a new condition which arises in the present derivation.
Note that
this is only a
condition for initial $H$ and $F$
because
$\tr\brak{H(t)F(t)}$
is constant in time
by
the
QB equation~\eqref{eq-QBeq}.
Moreover, because the overall scale of $F$ is arbitrary,
eq.~\eqref{eq-QB-surface}
essentially states that $\tr\brak{F(0)H(0)} \ne0$
and
merely gives of a normalization of $F$.

We will discuss
the role of the condition~\eqref{eq-QB-surface}
in Sec.~\ref{sec-drift-ineq} and \ref{sec-pmp-singular}
from the viewpoint of the maximum principle,
which leads to the
same condition~\eqref{eq-PMP-const}.
Here, we shall just see
how the condition~\eqref{eq-QB-surface}
was interpreted in
the previous studies~\cite{Car.Hos.Koi.Oku2006,
Car.Hos.Koi.Oku2007,Car.Hos.Koi.Oku2008,
Gen.Wu.Wan.Xu.Shi.Xie.Ron.Du2016b,
Wan.All.Jac.Llo.Lup.Moh2015}.
Consider
the Hamiltonian of the form
$H(t)=H_d+H_c(t)$ with
constraints
$\tr[H_c^2]=\Omega$
and
$H_c\in\CC$,
where $\Omega>0$ and $\CC$
is a certain subspace of $\su(N)$.
This is a 
common class of constraints
and was treated in
the previous studies. 
Then, eq.~\eqref{eq-QB-F} becomes
$F(t) = \lam(t)H_c(t)+F'(t)$.
When $H_d=0$,
the condition~\eqref{eq-QB-surface}
is equivalent to
$\lam\neq0$,
which was implicitly assumed 
in the previous studies.
The condition~\eqref{eq-QB-surface}
guarantees this implicit assumption.
On the other hand,
the case $H_d\neq0$ allows
the solutions with $\lam=0$
in general.
This correspond to
a singular control,
which will be
defined and discussed in Sections~\ref{sec-drift-ineq}
and
\ref{sec-pmp-singular}.

Note that
we can also consider
the time-optimal control problem
of \textit{state evolution} in QB
theory~\cite{Car.Hos.Koi.Oku2006}, which is the problem of finding
the Hamiltonian $H(t)$
that generates
the evolution of a quantum state
$\ket{\psi(t)}$
with specified quantum states at $t=0$ and $t=T$,
in the least possible time.
For this problem,
QB gives the same
conditions [eq.~\eqref{eq-QB-F},
\eqref{eq-QBeq}, and \eqref{eq-QB-surface}]
and
an additional condition
$F=FP+PF$, where
$P(t):=\ket{\psi}\!\bra{\psi}$.

\subsection{Derivation}
\label{qb-derivation}
Let us give a derivation of the result in the previous subsection.
We take variations of $S$ and
derive the Euler-Lagrange equations.

The variation with respect to $F(t)$
leads to the Schr\"odinger
equation~\eqref{eq-sch}.
The variation with respect to $\lam_j(t)$
leads to
the constraints
$f^j(H)=0,\;j=0,1,\dots,p$,
which is equivalent to eq.~\eqref{eq-qb-eqconstr}. 
The variation with respect to $H(t)$ yields
\begin{eqnarray}
  \delta S
  &=\int_0^T dt\;
  \tr\brak{
    \paren{-F(t) +\sum_j \lam_j(t) \f{\partial f^j(H)}{\partial H}}
    \delta H(t)}.
\end{eqnarray}
Since $\delta H(t)$ is an arbitrary
traceless Hermitian operator,
we obtain eq.~\eqref{eq-QB-F} that gives the form of $F$.

Let us take the variation with respect to
$U(t)$ and $T$.
We have
\begin{eqnarray}
  \delta S
  &=\int_0^T dt\;
  \tr\brak{
    iF(t)\dot{U}\delta U^\dag
    -i\dot{F}(t)\delta UU^\dag
    -iF(t)\delta U\dot{U}^\dag
  } \nonumber\\
  &\hspace{10em}
  +i\tr\brak{F(T)\delta U(T)U(T)^\dag}
  +\delta T,
  \label{eq-var-U}
\end{eqnarray}
where we have performed
integration by parts
and have used $\delta U(0)=0$.
We have also applied $L_S(t=T)=L_C(t=T)=0$
(after taking the variation),
which follow from the
constraints \eqref{eq-sch}
and \eqref{eq-qb-eqconstr}.
Using the Schr\"odinger equation \eqref{eq-sch}
and 
$\delta U^\dag = -U^\dag\delta UU^\dag$,
which follows from the unitarity of $U$,
we can rewrite the integrand in eq.~\eqref{eq-var-U} as
\begin{eqnarray}
  \tr\brak{\paren{-FH
      - i\dot{F}
      +HF
    }\delta UU^\dag}.
\end{eqnarray}
Since $\delta UU^\dag$ is
an arbitrary traceless
Hermitian operator,
$\d S=0$ implies the
QB equation~\eqref{eq-QBeq}.
The surface terms (i.e., non-integral terms) in
eq.~\eqref{eq-var-U}
become
\begin{eqnarray}
  \paren{-\tr\brak{H(T)F(T)}+1}\delta T,
\end{eqnarray}
because $\delta U=iHU\delta T$ holds at $t=T$
by eqs.~\eqref{eq-sch} and \eqref{eq-terminal-condition}.
This leads to the algebraic condition~\eqref{eq-QB-surface}
since $\tr\brak{HF}$ is constant by virtue of eq.~\eqref{eq-QBeq}.
We have thus shown the result of QB in the previous subsection.

\subsection{Some direct consequences}
\label{qb-direct-cons}
We shall see two important results
which can be
shown by QB immediately.

The first
consequence is
(see also \cite{Car.Hos.Koi.Oku2006}) that
\textit{the time-optimal
Hamiltonian is constant
in general
when the available Hamiltonians
have no restriction except
a normalization on its magnitude.}
Let $H(t)$ be subject to a unique constraint
\begin{eqnarray}
  0=f^0(H):= \frac12\tr\brak{H^2}-\Omega^2.
\end{eqnarray}
That is, $H(t)$ can be any
traceless
Hermitian operator satisfying
$\tr\brak{H^2}/2=\Omega^2$.\footnote{
  As was discussed in Sec.~\ref{sec-setup},
  the Hamiltonian
  $H(t)$ can always be considered as traceless without loss of generality
  (by considering the traceless part if not).
}
Then
the definition \eqref{eq-QB-F} of $F$ and
the QB equation \eqref{eq-QBeq}
yield
\begin{eqnarray}
  F(t) = \lam_0(t) H(t), \label{eq-nocons1}
  \quad
  i\dot{F}(t) = [H(t),F(t)], \label{eq-nocons2}
\end{eqnarray}
respectively.
From the latter, we see that $\tr\brak{F^2}$ is constant.
Then, squaring the former and taking the trace,
we have $\lambda_0(t)=$ constant.
Again from the latter, we have
\begin{eqnarray}
  \lam_0\dot{H}(t) = \lam_0[H(t),H(t)] = 0.
\end{eqnarray}
Therefore,
the Hamiltonian $H(t)$
is
constant
in time.
Geometrically, this is understood as
the evolution $U(t)=e^{-iH(0)t}$ is along a geodesic on
$\SU(N)$ when essentially no restriction is placed on $H(t)$.

The second consequence is
that
\textit{the QB equation of
the Zermelo navigation problem~\cite{Rus.Ste2014, Bro.Mei2015a,Rus.Ste2015}
reduces to
that of 
the above-discussed drift-free problem and is thereby
easily solved.}
Let the Hamiltonian $H(t)$ consist
of a fixed drift part $H_d$ and a controllable part $H_c(t)$,
\begin{eqnarray}
  H(t)=H_d+H_c(t).
  \label{eq-drift+cont}
\end{eqnarray}
In our formulation,
the Zermelo navigation problem is the case with a simple constraint
on $H_c$:
\begin{eqnarray}
  0=\tilde f^0(H_c):=\frac12\tr\brak{H_c^2}-\Omega^2,
\end{eqnarray}
This problem is easily solved by  moving to the interaction picture.
We define the interaction picture
operator $A_I(t)$
of a Hermitian Schr\"odinger picture operator $A(t)$ by
$A_I(t):=e^{iH_dt}A(t)e^{-iH_dt}$ and that of time evolution operator
$U(t)$ by $U_I(t):=e^{iH_dt}U(t)$.
We observe that
$\tilde f^0(H_{c,I})=\tilde f^0(H_c)$.
Because $f^0(H)=\tilde f^0(H-H_d)$,
eq.~\eqref{eq-QB-F}
yields
$F(t) = \lam_0(t) H_c(t)$.
Therefore, eqs.~\eqref{eq-sch}, \eqref{eq-QB-F} and \eqref{eq-QBeq}
are equivalent to
\begin{eqnarray}
  i\dot{U_I} = H_{c,I}U_I,
  \quad
  F_I = \lam_0 H_{c,I},
  \quad
  i\dot{F_I} = [H_{c,I},F_I].
\end{eqnarray}
These are nothing but eqs.~\eqref{eq-sch} and
\eqref{eq-nocons2},
with $H(t)$, $F(t)$ and $U(t)$
being replaced
by $H_{c,I}(t)$, $F_I(t)$ and $U_I(t)$.
Therefore, we immediately obtain the solution
$H_{c,I}(t)=H_{c,I}(0)=H_{c}(0)$ (constant) and
$U_I(t)=e^{-iH_c(0)t}$, that is,
\begin{eqnarray}
  H (t) = H_d+e^{-iH_dt}{H}_c(0)e^{iH_dt},
  \q
  U(t)=e^{-iH_dt}e^{-iH_c(0)t}.
  \label{eq-Qzermelo}
\end{eqnarray}
This is the same as the result
of Refs.~\cite{Bro.Mei2015a,Rus.Ste2015},
but the derivation is simpler and more
intuitive.
We have only appealed to the invariance of the constraint
$\tilde f^0(H_c)=0$ under transfer to the interaction picture.

This observation also provides a generalization of the consequence.
\textit{When the Hamiltonian $H(t)$ consists
of a drift $H_d$ and a controllable part $H_c(t)$ where
the constraints $\tilde f^j(H_c)=0$ are
invariant under transfer to the interaction picture (by $H_d$),
the QB equation reduces to that of a drift-free problem.}
Suppose that $\tilde f^j(H_c)=\tilde f^j(H_{c,I})$.
Then, we have
\begin{eqnarray}
  S\brak{H_d+H_c,U,F,\brac{\lambda^j};f^j}
  =
  S\brak{H_{c,I},U_I,F_I,\brac{\lambda^j};\tilde f^j},
  \label{eq-int-pict-action}
\end{eqnarray}
where the action functional $S\brak{H,U,F,\brac{\lambda^j};f^j}$
is given by
\eqref{eq-qb-action},
\eqref{eq-qb-LS} and
\eqref{eq-qb-LC}.
Note that the right hand side of
\eqref{eq-int-pict-action} is an action for a drift-free problem
with $H(t)=H_{c,I}(t)$.
Thus, the definition \eqref{eq-QB-F} of $F$ and
the QB equation \eqref{eq-QBeq}
for the original problem is equivalent to
those for a drift-free problem with constraints
\begin{eqnarray}
  \tilde f^j(H_{c,I})=0.
\end{eqnarray}
The solution to the original
problem is written as
\begin{eqnarray}
  H(t)=H_d+e^{-iH_dt}H_{c,I}(t)e^{iH_dt},
  \quad
  U(t)=e^{-iH_dt}\TT e^{-i\int_0^t ds H_{c,I}(s)},
\end{eqnarray}
where $\TT$ denotes the time-ordered product.
We remark that,
as for the algebraic equation,
we must use the original one \eqref{eq-QB-surface}
{} because
the condition
\eqref{eq-terminal-condition}
was for the fixed target unitary operator
$U_f$.

\subsection{Remarks}

The consequences in the previous subsection,
though they are very simple,
demonstrate that QB is useful in discussing
general features of time-optimal quantum control.
It is also useful in solving concrete time-optimality problems.

The task in QB is to solve
the QB equation~\eqref{eq-QBeq}
and Schr\"odinger equation~\eqref{eq-sch}
with boundary values
$U(0)=1$ and $U(T)=U_f$.
This is
a boundary value problem (BVP).
In some situations,
one can obtain analytical
solutions~\cite{Car.Hos.Koi.Oku2006,
Car.Hos.Koi.Oku2007,Car.Hos.Koi.Oku2008}.
However, more commonly,
one must employ some
numerical approach.
Numerical methods
for BVP include 
single or multiple shooting methods,
finite-difference methods,
and variational methods~\cite{Sto.Bul2002}.
These all convert a BVP
to the problem of finding roots of
a set of nonlinear equations
and solve them by numerical search
methods, such as Newton or
quasi-Newton methods.
To find good initial guesses,
one can for example make use of
a geometric algorithm~\cite{Wan.All.Jac.Llo.Lup.Moh2015},
which is similar in spirit to
homotopy methods~\cite{Sto.Bul2002}.
The algorithm
gradually
strengthens penalties on
the prohibited terms in the Hamiltonian.

Another way to obtain the time-optimal control
is to solve the
fidelity-optimal control problems
repeatedly, of which
the task is to
find the Hamiltonian $H(t)$
which maximizes the fidelity
of the terminal unitary operator
$U(T)$ and the target
$U_f$ for a fixed time $T$.
The time-optimal control
can be obtained
as
the fidelity-optimal control
with the minimal time
in which the
optimal fidelity attains
unity~\cite{Can.Mur.Cal.Faz.Mon.Gio.San2009}.
One can solve
the fidelity-optimal control problems
by the Krotov method~\cite{Pal.Kos2002,Pal.Kos2003},
gradient ascent pulse engineering (GRAPE)~\cite{Kha.Rei.Keh.Sch.Gla2005},
or chopped random basis (CRAB)~\cite{Can.Cal.Mon2011} for example.
QB can also be recast to
a fidelity-optimal control problem
and can be solved by a Krotov-type method~\cite{Koi.Oku2010}.
There, the same equations, the QB equation~\eqref{eq-QBeq}
and eq.~\eqref{eq-QB-F}, are solved
with a different boundary condition.

\section{Pontryagin's Maximum Principle
  for time-optimal control in quantum systems}
\label{sec-pmp}

Pontryagin's maximum
principle ({\MP}) is an optimal
control theory
which is applicable to control systems with
inequality constraints (e.g. \cite{Bol}).
There have been many
studies~\cite{Kir.Hoc.Spi2008,Boo2012,Rom2014,Alb.DA2015,Heg2013,Lap.Zha.Bra.Gla.Sug2010,Zha.Lap.Sug.Bra.Gla2011,Avi.Fis.Lon.Ger2014}
which make use of {\MP}.
In those studies,
individual quantum control problems
are
first transformed
to control systems
with concrete real variables,
e.g. the Bloch vectors or the Euler angles,
and then {\MP} is applied.
Such translation
depends on the physical system 
and individual control settings. 
Here, we prefer
to write down
the time-optimal control theory
based on {\MP}
in a general form,
as was done in QB.
This is suitable for general discussions 
as well as
for applications to various quantum systems.
This is also
a basis for the discussion in the subsequent sections.
In this paper,
we call the theory the generalized QB by {\MP}, or {\QMP}, for convenience.

\subsection{Formulation}
\label{formulation}

We
state the result of {\QMP} here and
put a proof in
Sec.~\ref{sec-app1}.

Consider the problem presented in Sec.~\ref{sec-setup}.
Let $H(t)$, where $0\le t\le T$, be
the time-optimal Hamiltonian
for given
$(U_f, \AA)$,
where 
$\AA$ is the set of available Hamiltonians and 
$U_f$ is the target unitary operator. 
Namely, $H(t)\in\AA$ drives the unitary operator $U(0)=1$ to
$U(T)=U_f$ through
the Schr\"odinger equation \eqref{eq-sch}
in the smallest time $T$.
Then, there exists a
Hermitian operator $F(t)\in \su(N)$
which satisfies
\begin{eqnarray}
  \tr\brak{KF(t)} &\le \tr\brak{H(t)F(t)},
  \label{eq-PMP-max}
\end{eqnarray}
for any
$K\in\AA$
at each $t$,
\begin{eqnarray}
  i\dot{F}(t) = [H(t),F(t)],
  \label{eq-PMP-QBeq}
\end{eqnarray}
and
\begin{eqnarray}
  \tr\brak{H(t)F(t)}=-p_0,
  \label{eq-PMP-const-normal-abnormal}
\end{eqnarray}
where $p_0\le0$.
A control protocol with
$p_0<0$ is called \textit{normal}
and that with $p_0=0$ is called \textit{abnormal}
(e.g. \cite{Bar.Mun2009}).
In this paper, we consider only
normal control protocols,
leaving the analysis of abnormal protocols
in {\QMP} for future work.
Thus, we reduce
eq.~\eqref{eq-PMP-const-normal-abnormal} to
\begin{eqnarray}
  \tr\brak{H(t)F(t)}=1,
  \label{eq-PMP-const}
\end{eqnarray}
by suitably rescaling the operator $F$.
We remark that the quantity
$\tr\brak{H(t)F(t)}$ is
constant in time, which follows from eq.~\eqref{eq-PMP-QBeq}.
To summarize, the time-optimal control problem
is transformed to the
problem of finding
a pair $\paren{H,F} \in (\AA,\su(N))$
satisfying
the conditions~\eqref{eq-PMP-max},
\eqref{eq-PMP-QBeq}, and \eqref{eq-PMP-const}.

In the theory of maximum principle,
$H_{\mathrm{{\MP}}}\paren{H(t),F(t)}=
-1+\tr\brak{H(t)F(t)}$ is called
the \textit{Pontryagin Hamiltonian}.
Then eq.~\eqref{eq-PMP-max}
states that the optimal control Hamiltonian
$H(t)$
is the maximizer of
the Pontryagin Hamiltonian 
at each $t$.

\subsection{Relation between QB and {\QMP}}

{\QMP} can be seen as
a generalization of QB
because
{\MP} can be seen as a generalization of
classical variational calculus.
Although we could also have
generalized QB to the case of inequality constraints
by other methods
such as
the Karush-Kuhn-Tucker conditions (e.g. \cite{Boy}),
we adopted {\MP} because of the following advantages.
{\MP} has already been considered
in the optimal control problems on
Lie groups (e.g. \cite{Spi2013})
and
there have been
a number of studies
on a singular control,
which is the subject of
Sections~\ref{sec-pmp-singular}.

{\QMP} has
the same equations~\eqref{eq-PMP-QBeq} and
\eqref{eq-PMP-const}
as QB does, namely
the QB equation \eqref{eq-QBeq} and eq.~\eqref{eq-QB-surface},
though
the definition of
the operator $F$
is different.
In {\QMP},
the operator $F$
is determined indirectly
by the maximum condition~\eqref{eq-PMP-max}.
In the case of equality constraints,
eq.~\eqref{eq-PMP-max} in {\QMP},
which determines $F(t)$,
reduces to
eq.~\eqref{eq-QB-F} in QB,
the definition of $F$.
To see
this,
let us
assume
that
the set $\AA$ of available
Hamiltonians is given by
eq.~\eqref{eq-qb-eqconstr}.
From eq.~\eqref{eq-PMP-max},
the time-optimal Hamiltonian
maximizes $-1+\tr\brak{H(t)F(t)}$ at any
time $t$
while satisfying the equality constraints
$f^j(H)=0$.
Then, by
the method of
Lagrange multipliers,
the function of $H$,
\begin{eqnarray}
  -1+\tr \brak{H(t)F(t)} -\sum_j\lambda_jf^j(H)
\end{eqnarray}
must have an extremum.
We obtain
$F = \sum_j\lambda_j\f{\partial f^j(H)}{\partial H}$
from the
condition that the derivative vanishes.
This is
nothing but eq.~\eqref{eq-QB-F}.
Thus, {\QMP} includes QB
as a special case where
constraints on $H(t)$ are expressed only by equalities.

We shall comment on the
difference between eq.~\eqref{eq-QB-surface}
and eq.~\eqref{eq-PMP-const}.
A strict application
of the method of Lagrange multipliers
also admits abnormal control protocols
(e.g. \cite{Mon1992} and \cite[Theorem 74.1]{Bli1946}),
though we did not consider such cases for simplicity
in our derivation of QB.
In such cases, eq.~\eqref{eq-QB-surface} in QB
will become $\tr\brak{HF}=-p_0$ with
Lagrange multiplier $p_0$.

\begin{figure}[b]
  \centering
  \includegraphics[width = 0.6\textwidth]{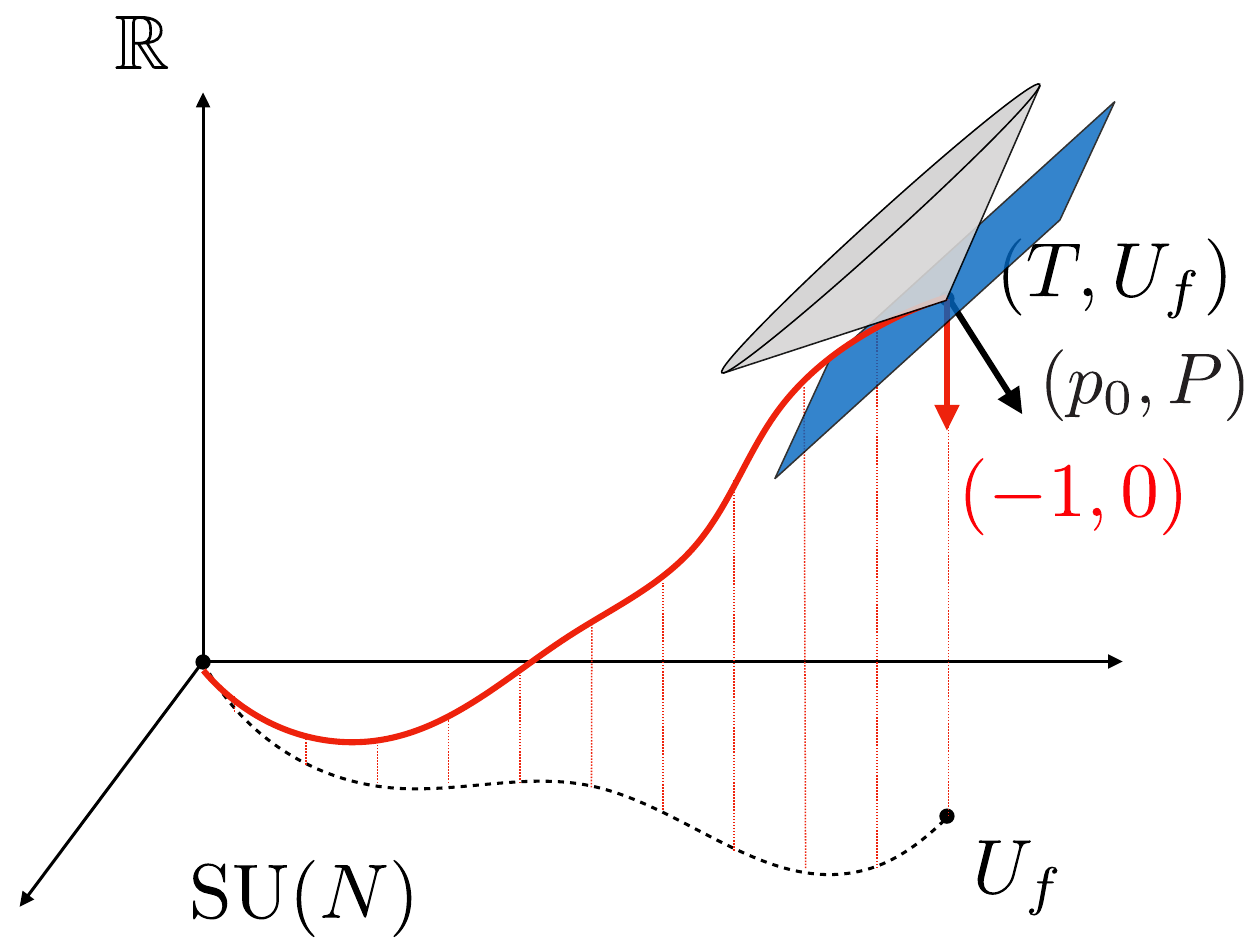}
  \caption{The augmented system
    $\mathcal{M}:=\mathbb{R}\times \SU(N)$
    and the trajectory of the optimal control.
    The
    horizontal plane
    depicts 
    the
    space
    $\SU(N)$ and the
    vertical axis
    is the
    extra dimension of time cost.
    The black dashed curve is a trajectory
    in the original quantum control system and the red bold curve
    represents the corresponding trajectory in the augmented system
    \eqref{eq-aug}.
    The gray cone is the tent
    $\tilde{\RR}_{(T,U(T))}$.
    The blue plane is
    the hyperplane that divides the tangent space at $(T,U(T))$
    into a half-space containing the tent
    and 
    that containing
    the vector $(-1,0)$.
    The vector $(p_0,P)$ is normal to the hyperplane.
  }
  \label{fig-pmp-aug}
\end{figure}

\subsection{Derivation}
\label{sec-app1}

We provide a derivation
of the statement of {\QMP} in
Sec.~\ref{formulation}.
Although
it is an adaptation of Pontryagin's {\MP}~(e.g. \cite{Bol}),
especially that
for control systems on a Lie group 
(e.g. \cite{Spi2013}),
to the problem of time-optimal quantum control,
it may clarify some subtle points of {\QMP} including
the difference from the conventional QB.
The derivation also reveals the origin of $F(t)$.

Consider an \textit{augmented system}
$\mathcal{M}:=\mathbb{R}\times \SU(N)$
(see Fig.~\ref{fig-pmp-aug}).
A trajectory $\paren{t,U(t)}$ in
$\mathcal{M}$ 
represents
an evolution of a unitary operator $U(t)$
and the time cost $t$ for each $U(t)$ in that evolution.

For a given
protocol
$H(t)$,
the trajectory in $\MM$
is determined by the differential equation
\begin{eqnarray}
  \dot{t} &= 1,\q
  \dot{U}(t) & = -iH(t) U(t).
  \label{eq-aug}
\end{eqnarray}
The \textit{reachability set} $\mathcal{R}$
in the augmented system is defined by
\begin{eqnarray}
  \mathcal{R} &:= \bigcup_{t\in[0,\infty]} \mathcal{R}_t,\q
  \mathcal{R}_t &:= \brac{(t,U(t))| H(t)\in\AA},
\end{eqnarray}
where
$\mathcal{R}_t$ represents
the set of all reachable points $U(t)$
at time $t$ and
$U(t)$ is generated from
the Schr\"odinger equation~\eqref{eq-sch}
with the initial condition $U(0)=1$.

Let the final time $T$ and the protocol
$H(t)$ be optimal for given $(U_f,\AA)$, so that $U(T)=U_f$.
Then, the reachability and the optimality imply that
the terminal point
$(T,U(T))$
of the trajectory $(t, U(t))$
lies at the boundary of $\RR$.

We would like to discuss the changes of the point
$(T,U(T))$
caused by the changes of
the protocol $H(t)$ and
the final time $T$.
For the former, we employ
the \textit{needle variation},
which is the variation of $H(t)$ on infinitesimal intervals
but allows finite changes of $H(t)$ there.
This is particularly useful
when the optimal protocols may have finite jumps, or
when the change of $H(t)\in\AA$ is allowed
only on
one side, i.e., when $\d H(t)$ is allowed but $-\d H(t)$ is not.
A simple needle variation
$M(\ta;\d\ta;K): H(t)\mapsto H'(t)$
at $t=\ta\in(0,T)$
where $H(t)$ is continuous
is defined by
\begin{eqnarray}
  H'(t)=\left\{
  \begin{array}{ll}
    K, &  \ta-\d\ta<t\le \ta, \\
    H(t), & 0\le t\le \ta-\d\ta \mbox{ and } \ta<t\le T,\\
  \end{array}
  \right.
  \label{eq-McShane-simple}
\end{eqnarray}
where
$\d\ta\ge0$ is infinitesimal
and $K\in\AA$.
By appropriately defining addition and nonnegative scalar
multiplication, 
which are essentially operations on the time intervals,
the needle variations form a space which is closed in those
operations,
though
the ``addition'' is noncommutative
(see \ref{sec-app2}).

The variation $H(t)\mapsto H'(t)$ with respect to $T$ is defined as follows.
If $\d T\le 0$,
$H'(t)$ is simply the
restriction of $H(t)$ on the shortened interval, i.e.,
$H'(t)=H(t)$,
$0\le t\le T-\d T$.
If $\d T> 0$,
$H'(t)$
is defined on the extended interval by the value at $t=T$, i.e.,
$H'(t)=H(t)$,
$0\le t\le T$ and
$H'(t)=H(T)$,
$t> T$.
It can be seen that the combination of the two kinds of variations
[of $H(t)$ and $T$]
above
again
form a space which is closed under addition and nonnegative scalar
multiplication.
Furthermore, though the ``addition'' is noncommutative,
they become commutative
in the resulting first-order variation of
the final unitary operator $U(T)$
(\ref{sec-app2}).
Thus,
$\d \brak{U(T)}$ form a convex cone contained in $T_{(T,U(T))}\MM$.
This cone is called
the \textit{tent}
$\tilde{\RR}_{(T,U(T))}$
of
${\RR}$
at
$(T,U(T))$.

If the tent
$\tilde{\RR}_{(T,U(T))}$ contains the downward vector $(-1,0)$
in its interior, the protocol $\paren{H(t),F(t)}$ cannot be optimal.
This follows from the fact that the deviated points from $(T,U(T))$
by the variations of $H(t)$ and $T$,
not only in the first order but to all orders,
form
a deformed cone in $\MM$ that intersects the segment
$[0,T)\times \brac{U_f}$.
Thus, if $(H(t),F(t))$ is
time optimal,
there is a hyperplane that
separates the tangent space
$T_{(T,U_f)}\mathcal{M}$
into a closed half-space containing the tent
and a closed half-space containing
the vector $(-1,0)$.
In terms of the normal vector to the hyperplane,
there exist a nonzero vector 
$(p_0,P)\in T_{(T,U(T))}\MM$
such that\footnote{
  We define the tangent vector $P$ at $U\in \SU(N)$ 
  as a Hermitian operator 
  such that the change of $U$ caused by $\epsilon P$, 
  where $\epsilon\in\mathbb{R}$ is infinitesimal, 
  is $-i\epsilon PU$.
} 
\begin{eqnarray}
  \angl{(p_0,P),\; (-1,0)}\ge0\ge
  \angl{(p_0,P),\; (\d T,\, i \d \brak{U(T)} U(T)^\dag)}, 
  \label{eq-separation}
\end{eqnarray}
where $\d \brak{U(T)}$ is the first-order variation
and
the bracket denotes the inner product
$\angl{(q_0,Q),(r_0,R)}:=q_0r_0+\tr\brak{QR}$.
The first inequality in \eqref{eq-separation} 
immediately implies $p_0\le0$.

The first-order variation of $U(T)$
caused by $M(\ta;\d\ta;K)$ in
\eqref{eq-McShane-simple} is
\begin{eqnarray}
  \delta U(T)
  =
  U(T,\tau)\paren{-i\delta \tau\brak{K-H(\tau)} }U(\tau),
  \label{eq-dU-dH}
\end{eqnarray}
where $U(t,\tau)$ is the
time evolution operator
satisfying the
the Schr\"odinger equation~\eqref{eq-sch}
with the initial condition $U(\tau,\tau)=1$.
Let us define a Hermitian operator 
\begin{eqnarray}
  F(t):=U(t,T)PU(T,t). 
\end{eqnarray}
It is immediately seen that $F$ satisfies the QB equation
\eqref{eq-PMP-QBeq}. 
From \eqref{eq-separation} and
\eqref{eq-dU-dH}, we also have
\begin{eqnarray}
  \tr\brak{KF(\tau)} \le \tr\brak{H(\tau)F(\tau)}.
\end{eqnarray}
Because $\ta$ is arbitrary, this is nothing but
the maximum condition \eqref{eq-PMP-max}.

The first-order variation of $U(T)$ with respect to $T$ is
given by
\begin{eqnarray}
  \delta \brak{U(T)} = (-i\delta T) {H(T)} U(T).
  \label{eq-dU-dT}
\end{eqnarray}
From \eqref{eq-separation} and
\eqref{eq-dU-dT}, and because $\d T$ can be of both signs,
we have
\begin{eqnarray}
  \tr\brak{H(T)F(T)} = -p_0 \ge0.
\end{eqnarray}
This is the algebraic condition~\eqref{eq-PMP-const}.

\section{Drift and inequality constraints}
\label{sec-drift-ineq}

We have
discussed {\QMP},
a general
time-optimal quantum control theory
for arbitrary constraints $\AA$, where $H(t)\in\AA$.
For further analysis on the time-optimal control, 
we discuss the basic structure of $\AA$ in this section.
We
show
the conditions
under which
an inequality constraint
can be reduced to an equality one.
The use of {\QMP} is essential in the following discussion.

\subsection{Drift and control Hamiltonians}
\label{dr-and-ctl}
Let  $\AA\subset\su(N)$
be a general constraint (set of available Hamiltonians).
Consider the smallest hyperplane that contains $\AA$, 
which we call \emph{control hyperplane}.\footnote{
  Mathematically, 
  the control hyperplane 
  is the intersection of all hyperplanes of
  arbitrary dimensions that contain $\AA$.
}
Take an arbitrary fixed Hamiltonian
$H_d\in\AA$ and view the hyperplane
as a linear subspace $\CC$ of $\su(N)$ whose origin is $H_d$.
Then,
we can write
$H(t)\in\AA$
as
\begin{eqnarray}
  H(t) = H_d + H_c(t),
  \q
  H_c(t)\in \AA-H_d\;(\subset\CC),
\end{eqnarray}
where $H_d$ is time independent.
We call the constraint $H_c(t)\in \CC$
a \textit{subspace constraint} and
$\CC$ the \textit{control subspace}.
We call $H_d$ and $H_c(t)$
the \textit{drift} and \textit{control}
Hamiltonians, respectively.
Physically,
the drift Hamiltonian $H_d$ describes
the fields intrinsic to the system
such as
an
interaction between the particles with a fixed coupling
or
a fixed magnetic
field.
The control Hamiltonian $H_c(t)$
describes the fields
that we can control, such as a pulse sequence of electromagnetic
waves
or an adjustable magnetic field.

A subspace constraint can be expressed by equality constraints
in the following simple manner.
Since the space $\su(N)$ can be written as
\begin{eqnarray}
  \su(N)=\CC\oplus\CC^\perp,
\end{eqnarray}
where
$\CC^\perp$
is
the orthogonal complement
of $\CC$ (with respect to the inner product $(1/2)\tr AB$),
a subspace constraint $H_c(t)\in\CC$
is written as
\begin{eqnarray}
  \tr\brak{H_c(t)\tau_j}=0, \quad j=\dim\CC+1,\dots,N^2-1,
\end{eqnarray}
where
an orthonormal basis
$\brac{\tau_j}_{j=1}^{N^2-1}$
on $\su(N)$
is chosen so that
$\tau_1,...,\tau_{\dim \CC}\in\CC$
and
$\tau_{\dim \CC+1},...,\tau_{N^2-1}\in\CC^\perp$.
The total
Hamiltonian is written as
\begin{eqnarray}
  H(t) = H_d + \sum_{j=1}^{l} u^j(t) \tau_j,
  \label{eq-typical-form}
\end{eqnarray}
with some real variables $u^j(t)$.

\begin{figure}
  \centering
  \includegraphics[width=\textwidth]{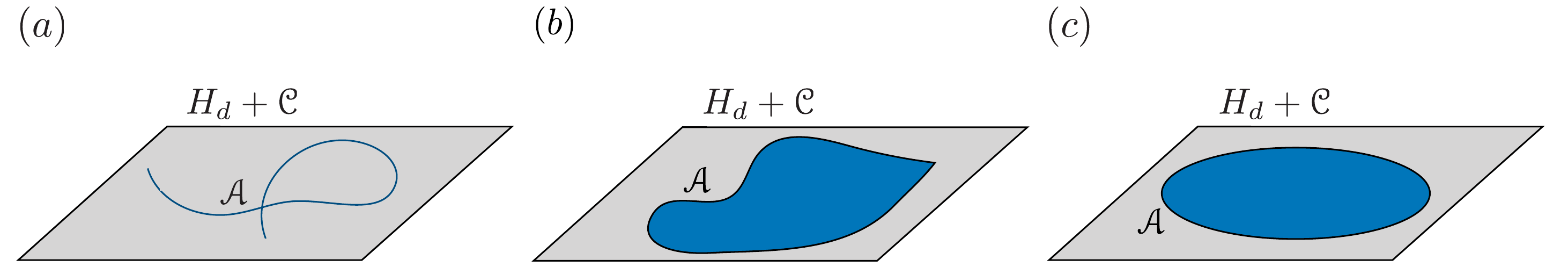}
  \caption{
    Illustration of
  (a) general, (b) planar, and
  (c) typical constraints. Each gray (or bright) plane
  represents the control hyperplane $H_d+\CC$ 
    and each blue (or dark)
  region represents the set $\AA$ of all available Hamiltonians.}
  \label{fig-gen-pla-typ}
\end{figure}

We shall define a theoretically convenient and 
practically common class of constraints. 
We say that
a constraint $\AA$ is \textit{planar}
if $\AA$ is a closed region in $H_d+\CC$ and $\dim\AA=\dim\CC$ holds, 
more precisely, 
if 
\begin{eqnarray}
\AA=\mathrm{closure}\paren{\mathrm{interior}(\AA)} 
\end{eqnarray}
in the topology of the hyperplane $H_d+\CC$ 
(see Fig.~\ref{fig-gen-pla-typ}). 
We also define a further restricted class.
We say that a constraint $\AA$ is \textit{typical}
if $\AA$ is defined by 
a subspace constraint $H_c(t)\in \CC$ 
and a single inequality 
\begin{eqnarray}
  \f12\tr\brak{H_c(t)^2} &\le \Omega^2, 
  \label{eq-amplitude}
\end{eqnarray}
where $\Omega>0$. 
The constraint \eqref{eq-amplitude} is 
called the 
\textit{finite energy constraint} and 
physically means that
the system has a finite energy bandwidth.
Thus, $\Omega(>0)$ gives an upper bound of
the (Hilbert-Schmidt) norm of the control Hamiltonian.
We can write $\AA$ of a typical constraint as
\begin{eqnarray}
  \AA
  & = & \brac{
    H_d+H_c \; \middle| \;
    \tr\brak{H_c\tau_j}=0, \; j>\dim\CC,\; \;
    \f12\tr\brak{H_c^2}\le \Omega^2
  },
\end{eqnarray}
and
the total
Hamiltonian as
eq.~\eqref{eq-typical-form}
with a constraint $\sum_j (u^j)^2\le\Omega$
(see Fig.~\ref{fig-gen-pla-typ}).

\subsection{Lollipop-type constraints allow reduction
to equality constraints}
\label{sec-lollipop}

\begin{figure}[t]
  \centering
  \includegraphics[width=0.6\textwidth]{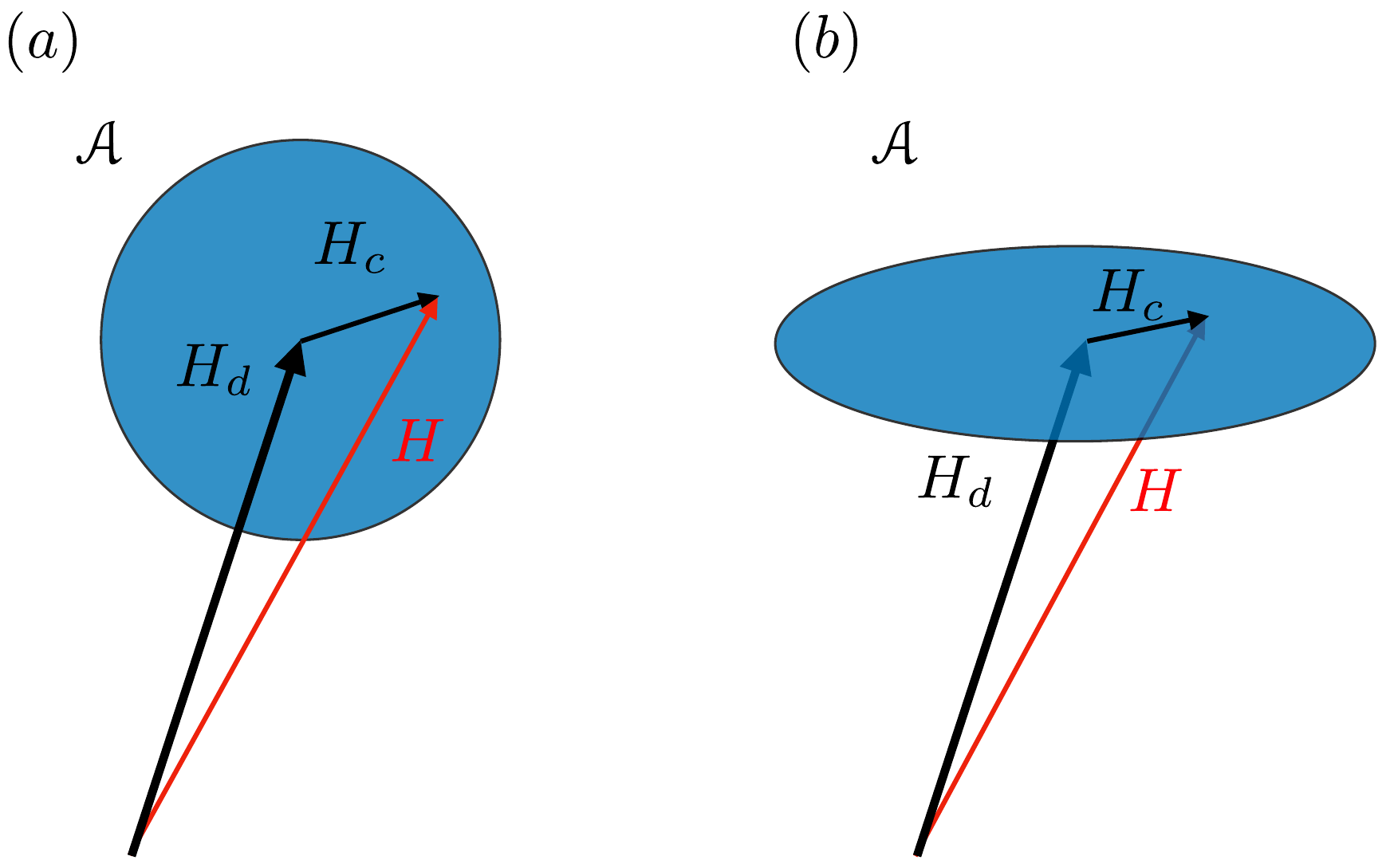}
  \caption{(a)
    An example of  ``lollipop-type'' constraint for $H$.
    This is characterized by $H_d\in\CC$, 
    where $H_d$ is the drift Hamiltonian and 
    $\CC$ is the control subspace.
    We can ``stretch'' the total Hamiltonian $H$ to the
    boundary of $\AA$ while keeping its direction.
    (b)
    An example of ``lotus leaf-type'' constraint, characterized by
    $H_d\not\in\CC$.
    We cannot enlarge the total Hamiltonian $H$
    without changing its direction.
  }
  \label{fig-lollipop-lotus}
\end{figure}

We divide the general (i.e., not necessarily planar) 
constraints into two types,
whether the drift $H_d$ is in the control
subspace $\CC$ or not.
We call the constraint (or $\AA$)
\textit{lollipop type} if
$H_d\in\CC$ and
\textit{lotus leaf type} if $H_d\not\in\CC$
(see Fig.~\ref{fig-lollipop-lotus}).
For simplicity and concreteness of the presentation,
we first discuss the {typical} constraints $\AA$ and then 
argue the results apply to the planar constraints as well.

If $\AA$ is typical and lollipop type,
including the case $H_d=0$,
the inequality constraint \eqref{eq-amplitude}
can be reduced to an equality
condition,
\begin{eqnarray}
  \f12\tr\brak{H_c(t)^2} = \Omega^2.
  \label{eq-amplitude-eq}
\end{eqnarray}
We give a brief proof.
In general, at any instant of time $t$, if we can stretch
$H(t)\in\AA$ to $a H(t)\in\AA$ with $a>1$, we can obtain a faster protocol
for the same unitary evolution $U_f$.
Thus, the time-optimal protocol must not allow
such a stretching $a>1$ at any time $t$.
As shown in
Fig.~\ref{fig-lollipop-lotus},
any $H(t)$ in the interior of $\AA$ allow such
a stretching.
Thus, the time-optimal protocol $H(t)$
must belong to the boundary of $\AA$
and
satisfy the equality \Ref{eq-amplitude-eq}
at any time $t$.
See also the supplemental material
of Ref.~\cite{Wan.All.Jac.Llo.Lup.Moh2015}.
For planar, lollipop-type  constraints,
the time-optimal control $H(t)$
belongs to the boundary of
$\AA$ in the hyperplane $H_d+\CC$.
Thus, we can reduce one of
the inequalities into the equality,
though we may not know which
equality is attained.
We remark that
for non-planar constraints, or $\dim\AA<\dim\CC$,
$H(t)$ always is regarded
as being in the boundary of $\AA$.

\subsection{Lotus-leaf-type constraints and a sufficient condition for
    the reduction}
  \label{sec-lotus-leaf}

As was defined in the previous subsection,
a general constraint $\AA$ is called lotus-leaf-type if $H_d\notin\CC$.
For such a constraint,
the situation differs from the lollipop-type case. 

The reduction of an inequality constraint to
an equality constraint
by
the stretching $aH(t)$ with $a>1$
is not possible in general.
We must consider the inequality
constraint~\eqref{eq-amplitude} in detail.
This can be seen by a simple example,
\begin{eqnarray}
 H(t) = \omega_0 \s^z + u(t) \s^x,
\end{eqnarray}
for a qubit,
where
$u(t)$ is a single control variable
with $|u(t)|\le\Omega$~\cite{Heg2013}.
In this case, the control subspace
is $\CC = \mathbb{R}\s^x$ and
is lotus leaf type.
When the target is
$U_f = e^{-i(\pi/4)\s^z}$,
the time-optimal control
is apparently $H(t) = \omega_0\s^z$,
hence $u(t) = 0$.
Thus, the problem is not solved
with an equality constraint
such as $|u(t)|=\Omega$.
It follows that, in general,
the inequality constraint
is essential for time-optimal
control under
lotus-leaf-type constraints.

\begin{figure}[t]
  \centering
  \includegraphics[width=0.3\textwidth]{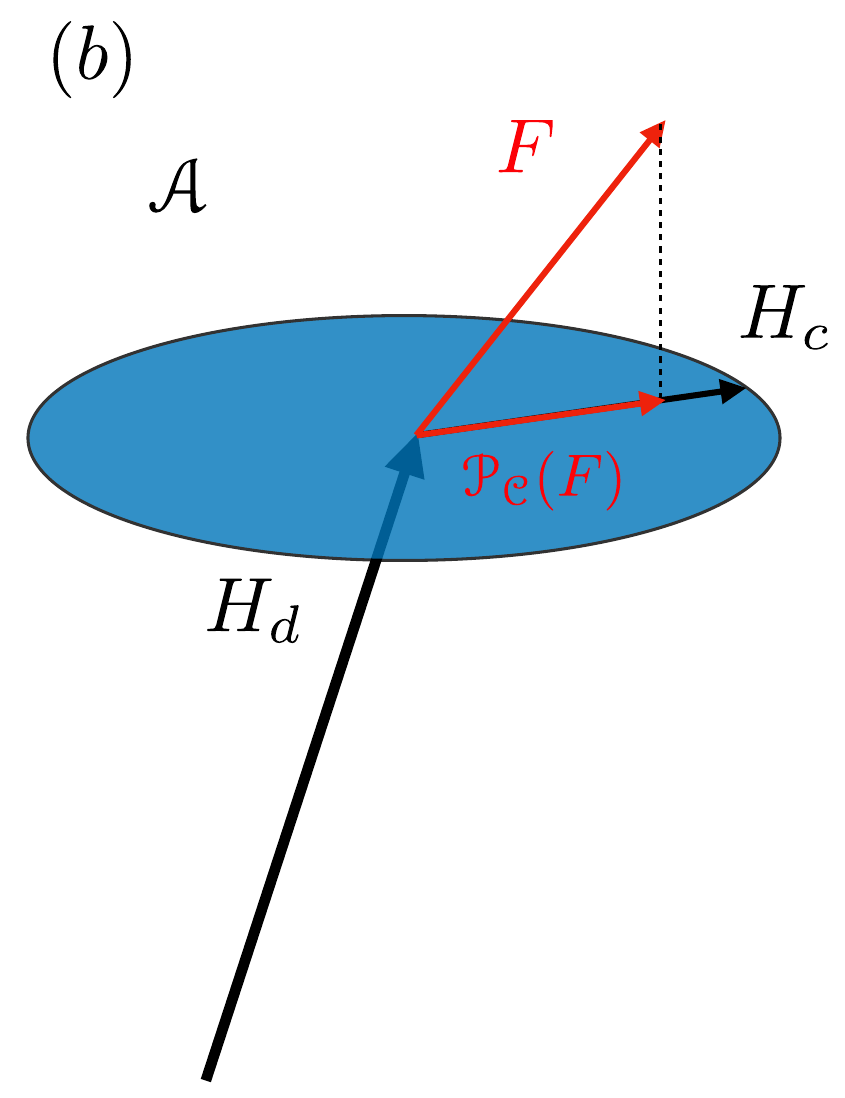}
  \caption{ The time-optimal control $H$ in
 a ``lotus-leaf-type'' constraint.
 If we have $\mathcal{P}_\CC(F)\neq0$,
 where $\mathcal{P}_\CC$ is a projection onto the subspace $\CC$,
 then the inner product $(1/2)\tr\brak{FH_c}$ has a maximum
 when $H_c$ is proportional to $\mathcal{P}_\CC(F)$
 with maximal norm.
}
  \label{fig-proof-max}
\end{figure}

However, 
we observe by {\QMP} that 
if an additional condition holds,
the inequality
constraint~\eqref{eq-amplitude}
can be reduced to an
equality one~\eqref{eq-amplitude-eq}
for
a
lotus-leaf-type
constraint $\AA$ as well.
Again, we first discuss the case of typical $\AA$ for simplicity and
see later that the results hold for planar constraints. 
Let $\AA$ be typical.
{\QMP} requires the existence of
a Hermitian operator $F$ which satisfies
the QB equation~\eqref{eq-PMP-QBeq}.
We assume that
$F$ has a nonzero projection onto
the subspace $\CC$,
or $\mathcal{P}_\CC(F)\neq0$,
where $\mathcal{P}_\CC$ is an
orthogonal
projection onto $\CC$ as in Fig.~\ref{fig-proof-max}.
Since {\QMP} requires the maximization
of the Pontryagin Hamiltonian
$-1+\tr\brak{H_dF}+\tr\brak{H_cF}$,
the time-optimal Hamiltonian
$H$ must maximize
the inner product
$(1/2)\tr\brak{H_cF}$.
By the Cauchy-Schwartz inequality,
we obtain
\begin{eqnarray}
 \tr\brak{H_cF} \le
  \sqrt{\tr\brak{H_c^2}\tr\brak{\mathcal{P}_\CC(F)^2}}
  \le \Omega
  \sqrt{2\tr\brak{\mathcal{P}_\CC(F)^2}}.
  \label{eq-cs-ineq}
\end{eqnarray}
This equality is attained
if and only if $H_c$ is proportional to $\mathcal{P}_\CC(F)$
and $H_c$ has the maximum norm
$\Omega$ (see Fig.~\ref{fig-proof-max}).
Therefore, the time-optimal
Hamiltonian $H$
must satisfy
the equality constraint~\eqref{eq-amplitude-eq}.
Note that the
``only if'' part is guaranteed by the
assumption $\mathcal{P}_\CC(F)\neq0$.

For a planar
constraint $\AA$,
we can still show that the time-optimal
$H$ belongs to the boundary of $\AA$,
with $\AA$ being considered
as a subset of
the plane
$H_d +\CC$.
If $H_c$ is in the interior
of $\AA-H_d(\subset\CC)$,
we can take a
nonzero variation $\delta H_c$
such that
$\tr\brak{\delta H_cF}>0$
(we still assume $\mathcal{P}_\CC(F)\neq0$).
Then we can make $\tr\brak{H_cF}$
larger by changing
$H_c$ to $H_c+\delta H_c$.
Note that
the assumption
$\mathcal{P}_\CC(F)\neq0$
guarantees the strict positivity
of $\tr\brak{\delta H_c F}$.
If we say
$\mathcal{P}_\CC(F)\neq0$
the regularity condition,
which is a term introduced in the next section,
the result is stated as follows.
\textit{If the constraint $\AA$ is planar,
  all regular optimal protocols are on the boundary of
  $\AA$.}

We have shown by 
{\QMP} that an inequality constraint reduces 
to an equality one even for the lotus-leaf-type
constraint 
with the help of the assumption
$\mathcal{P}_\CC(F)\neq0$.
Conversely, if this does not hold,
the inner product $(1/2)\tr\brak{H_cF}$
vanishes
for all $H_c$
so that 
the maximization
condition~\eqref{eq-PMP-max}
gives no information on the
optimal Hamiltonian.
Such
solutions to {\QMP}
are called singular,
which will be discussed in detail in the subsequent section.

\section{Singular controls and
  the generalized Legendre-Clebsch condition}
\label{sec-pmp-singular}

In this section,
we shall discuss
{singular} controls.
We showed by an example in the previous section that
singular controls
can be time optimal
if $\AA$ is lotus leaf type.
Inequality constraints cannot be reduced to equality ones 
for a singular control.
We will also see that {\MP} is not powerful enough to
determine the optimal protocol from singular ones.
Thus, it is desirable to restrict or exclude the possibility of
singular protocols to be optimal.
We shall give a precise definition of singular protocols 
in Sec.~\ref{sec-sing-def} 
and derive conditions beyond {\MP} that the singular optimal
protocols must satisfy 
in Sec.~\ref{sec-glc}. 
We discuss the physical meaning of singular controls in
Sec.~\ref{sec-sing-meaning}. 
As before,
the problem and the formulation
are the same as in Sec.~\ref{sec-setup} and
Sec.~\ref{formulation}.

\subsection{Singular controls}
\label{sec-sing-def}

Let $\AA$ be a general
constraint.
We say that
a quantum control protocol $\paren{H(t),F(t)}$ is
\textit{singular} at time $t$ if
the maximum condition~\eqref{eq-PMP-max} is
trivially satisfied for any $H(t)\in\AA$,
namely,
$\tr\brak{ KF(t) }=$ constant for all $K\in\AA$.\footnote{
  The definition of singularity may be slightly different 
  in some
  literature. See the final remark in this subsection.
}
In terms of $\CC$ and $H_d$ in Sec.~\ref{dr-and-ctl},
which are determined by $\AA$,
the singularity condition is equivalent to
\begin{eqnarray}
  \tr\brak{\CC F(t)} &= 0, 
  \label{eq-singular-control}
\end{eqnarray}
or $F(t)\perp \CC$ 
(Fig.~\ref{fig-proof-max}).
We call a control
$\paren{H(t),F(t)}$ \emph{regular} at $t$
if it is not singular.
If the optimal control $(H(t),F(t))$ is singular
in a certain finite time interval containing $t$, 
the time derivatives of any order of
eq.~\eqref{eq-singular-control} must hold at $t$,
that is,
\begin{eqnarray}
 \f{\mathrm{d}^n}{\mathrm{d}t^n}
 \tr\brak{\CC F(t)} =0,
 \label{eq-singular-control-dt}
\end{eqnarray}
for $n=1,2,\dots$.
For $n=1$,
we have
\begin{eqnarray}
  \tr\brak{\brak{\CC,H_d}F(t)} = 0.
  \label{eq-singular-control-dt1}
\end{eqnarray}
It follows from 
the algebraic condition~\eqref{eq-PMP-const} of {\QMP} 
and the singularity condition \eqref{eq-singular-control} that 
any optimal singular control
$\paren{H(t),F(t)}$
satisfies
\begin{eqnarray}
  \tr\brak {H_dF(t)}&=1.
  \label{eq-singular-control2}
\end{eqnarray}
This in particular implies that
\textit{if 
the constraint $\AA$ (which is not necessarily planar) is 
lollipop type, 
time-optimal singular controls do not exist.}
This is so 
because $H_d\not\in\CC$ is necessary for satisfying 
eqs.~\eqref{eq-singular-control} 
and 
\eqref{eq-singular-control-dt1}
at the same time. 

When singular controls exist,
there are two difficulties in finding the time-optimal control.
First, {\QMP} admits any protocol which is a
sequence of regular and singular controls
as candidates of the optimal
one,
so that we do not know in advance
how to determine the number and sequence of them.
Second, we cannot identify
the optimal singular control
by {\QMP} in general.
This is because $\hpmp$ no longer
depends on the control Hamiltonian $H_c(t)$
for singular controls
and we cannot obtain any
information from the maximum
condition~\eqref{eq-PMP-max}.
For the latter problem, however,
we will find an additional condition
for singular controls to be optimal in
Sec.~\ref{sec-glc}.

We shall make
some comments here
on
the definition of singular controls.
A common definition of a
singular control is as follows.
Let $\bm{u}\in \mathcal{U}$
be a vector of control variables
describing the available Hamiltonian $H\in\AA$.
Then a control protocol is
called singular
if there exists a variation of
control variables $\delta \bm{u}$,
by which
the variation of
the Pontryagin Hamiltonian
$\delta \hpmp$ vanishes
up to the second order (e.g. \cite{Rob1967}).
In this paper,
the control protocol
is called singular
if the Pontryagin Hamiltonian
is constant on $\AA$.
This definition of singularity
is narrowed from
the common one
in the following two points.
First, our definition requires
the variation vanishes to all orders.
Second, 
the variation $\delta \hpmp$
of the Pontryagin Hamiltonian
must vanish for $\delta \bm{u}$
in all directions. 
In particular, we call a protocol regular 
in the case that 
some of the control variables are determined by the
maximum condition and some are not.
Note that, in such cases,
we can
redefine $\AA$
as the subset of the original $\AA$ that maximizes $\hpmp$
and regard the controls as singular ones~\cite{Rob1967}.

\subsection{Generalized Legendre-Clebsch condition}
\label{sec-glc}

We shall derive an additional
necessary condition for a
singular control
to be optimal,
which is one of the main results of this paper.
The condition can exclude some of the singular protocols from 
candidates of the optimal ones and
is useful 
in determining 
the optimal protocol.

Let $H(t)$ be parametrized
by
control variables $u^j(t)$, $1\le j\le l$,
in a certain subset $\UU\subset\mathbb{R}^{l}$,
where we have used up all equality constraints so that
all $u^j$ are independent.
In the following,
we assume that
the control variables
$u^j$ are in the interior of
$\UU$
so that we can take
an arbitrary
infinitesimal variation
of the control variables
$u^j(t)\rightarrow u^j(t)+\delta u^j(t)$
at time $t$.
Even if the control variables $u^j$
are on the boundary of $\UU$, 
we still can regard
$u^j$ to be in the interior
of the boundary of
$\UU$ 
by defining new control
variables $u'^j$ with reduced
dimensionality,
thanks to 
the implicit function theorem.
We can then apply the same arguments below
by replacing $u^j(t)$ with $u'^j(t)$ 
and
redefining $l$ as the dimension of $u'^j$.
We will demonstrate the calculations of
such a case
in Sec.~\ref{sec-example} (and in \ref{sec-app3}).

Let us introduce
the \textit{generalized
  Legendre-Clebsch} (GLC) condition~\cite{Rob1967}.
The GLC condition comes from positive 
semidefiniteness of the second-order variation of 
the cost functional $T$
and is useful in determining 
the singular optimal protocols,
where MP becomes trivial.
We define an $l\times l$ matrix
$Q^{(m)}$, $m=0,1,2\dots$,  whose $(i,j)$ element is given by
\begin{eqnarray}
  Q^{(m)}_{ij}
  :=
   \pdiff{u^i}\brak{\paren{\diff{t}}^m
     \paren{\pdiff{u^j}
       \hpmp\paren{H(t),F(t)}}
   }.
  \label{eq-GLC-mat}
\end{eqnarray}
We denote by $M$
the smallest value of $m$
for which $Q^{(m)}$ has
at least one nonzero element.
Any optimal control
must satisfy the following two
conditions:\footnote{
  Our definition of singularity is slightly stronger than
  that in Ref.~\cite{Rob1967} so that
  the theorem there implies the assertion here.
}
\begin{enumerate}
\item The integer $M$ must be even.
\item If $M=2k$, then $(-1)^k Q^{(2k)}$ must be
  negative semidefinite.
\end{enumerate}
When $M=0$, these yield
$Q^{(0)}_{ij} \le0$
which is implied by MP.
When the protocol is singular,
$Q^{(0)}_{ij} =0$ and $M>0$ follow.
It can also be shown that
the matrix $Q^{(m)}$
is symmetric if $m$ is even
and anti-symmetric if $m$ is odd.
We remark that
the GLC condition
does not depend on
individual representations;
although the matrix
$Q^{(m)}$
does,
the conditions $Q^{(M)}=0$,
$Q^{(M)}\ge0$, and $Q^{(M)}\le0$
do not.
See \ref{app-coordi-trans} for a proof.

We shall go back to time-optimal quantum control. 
We provide the expressions of
$Q^{(m)}_{ij}$ for singular protocols
so that we can examine
the GLC condition in a step-by-step manner.
A derivation
is given in \ref{app-Q}.
The singularity condition 
implies
$Q^{(0)}_{ij}=
\paren{\partial^2/\partial u^i\partial u^j}\tr\brak{HF}=0$.
Assume that $Q^{(k)}=0$
hold for $k=0,1,\dots,m-1$.
Then, $Q^{(m)}_{ij}$ is given by
\begin{eqnarray}
  Q^{(m)}_{ij}
  &=
  -i\tr\brak{
  \brak{h_i,F}
  R^{(m-1)}_{j} },
  \label{eq-GLC-mat-tau}
  \end{eqnarray}
where
$h_i:=\partial H/\partial u^i$
and
$R^{(m)}_j$ is obtained by
the following recurrence relation,
\begin{eqnarray}
  R^{(m)}_{j} =
  \diff{t}R^{(m-1)}_{j}
  -i[R^{(m-1)}_{j},H],
  \q
  R^{(0)}_j = h_j.
  \label{eq-GLC-mat-tau-R}
\end{eqnarray}
In fact,
$R^{(m)}_{j}$ is
a quantity defined through $\hpmp$ as
$R^{(m)}_{j} :=
  \pdiff{F}
  \paren{\diff{t}}^{m}
  \pdiff{u^j} \hpmp$.
We can perform the GLC test in the
time-optimal quantum control problem
as follows:
\begin{enumerate}
\item[(o)] Let $m=0$.  We have
  $Q^{(m)}=0$.
\item
  \label{i-1}
  Increase $m$ by one and calculate
  $Q^{(m)}$
  by eqs.~\eqref{eq-GLC-mat-tau}
  and
  \eqref{eq-GLC-mat-tau-R}.
\item If $Q^{(m)}$ is identically zero, with the help of
  the previously obtained conditions,
  go to \eqref{i-1}.
\item If $m$ is odd, impose a condition $Q^{(m)}=0$
  and
  go to \eqref{i-1}.
\item Impose a condition $(-1)^kQ^{(2k)}\le0$
  and halt.
\end{enumerate}

In the special case of planar constraint $\AA$,
the Hamiltonian $H(t)$ can be written
by
independent control
variables $\brac{u^j}\in\UU$ as
\begin{eqnarray}
   H(t) = H_d + H_c(t) = H_d + \sum_{j=1}^{l} u^j(t) h_j.
   \label{eq-rep-control-Hamil}
\end{eqnarray}
Then, $h_j$ appearing in eqs.~\eqref{eq-GLC-mat-tau} and
\eqref{eq-GLC-mat-tau-R} become fixed, time-independent
operators in $\su(N)$,
which span the control subspace $\CC$.
We give concrete expressions for
the first few $Q^{(m)}_{ij}$ in this case:
\begin{eqnarray}
  Q_{ij}^{(1)} &= -i\tr\brak{\brak{h_j,h_i}F},\nn
  Q_{ij}^{(2)} &= \tr\brak{\brak{\brak{H,h_j},h_i}F},\nn
  Q_{ij}^{(3)} &= i\tr\brak{\brak{\brak{H,\brak{H,h_j}},h_i}F}
  + \tr\brak{\brak{\brak{\diff[H]{t},h_j},h_i}F}.
  \label{eq-GLC-mat-tau-concrete}
\end{eqnarray}
For example, the GLC condition for $m=1$ reads
$
  \tr\brak{\brak{h_i,h_j}F}=0
$
for
$  1\le i,j \le l
$. This is equivalent to
\begin{eqnarray}
  \tr\brak{[\CC,\CC]F}=0.
  \label{eq-GLC-1}
\end{eqnarray}
From this condition,
we observe that
\textit{time-optimal singular controls exist
only when $H_d\not\in[\CC,\CC]$.}
For
otherwise,
eq.~\eqref{eq-GLC-1}
would imply
$\tr\brak{H_dF}=0$,
which would contradict
eq.~\eqref{eq-singular-control2}.

\subsection{Physical meaning of singular controls}
\label{sec-sing-meaning}

In considering
the physical meaning of the singular controls, 
we naively
expect that the regular control takes
advantage of the control Hamiltonian
$H_c$ and
the singular control
takes advantage
of the drift Hamiltonian $H_d$.
This naive expectation is
supported by
the following facts.
(i) For typical $\AA$, the regular optimal controls
necessarily attain the maximal norm
of the control Hamiltonian;
for planar $\AA$, 
the regular optimal controls
must be on the boundary of $\AA$.
(ii)
For planar $\AA$,
optimal singular controls exist
only when
the drift Hamiltonian
$H_d$
satisfies
the conditions
$H_d\not\in\CC$
and $H_d\not\in[\CC,\CC]$.
The latter two conditions
imply that
$H_c(t)$ cannot
help generate
unitary operator
$e^{-i\a H_d}$
up to the second-order
of infinitesimal
time interval $\delta t$.
This follows
from the Baker-Campbell-Hausdorff formula.
For example,
we assume
$H(t) = H_d+K_1$
for $t\in[0,\delta t]$
and $H(t) = H_d+K_2$
for $t\in[\delta t,2\delta t]$,
where $K_1,K_2\in\CC$ and $H_d\not\in\CC,[\CC,\CC]$.
Then, we have
the resulting unitary operator
$U(2\delta t)$ as
\begin{eqnarray}
&\hspace{-3em}U(2\delta t)=
  e^{-i\delta t(H_d+K_2)}
  e^{-i\delta t(H_d+K_1)} \nn
  &=
  \exp\brak{-2i\delta tH_d + \delta t(K_1+K_2)
  +\f{\delta t^2}{2}
  \brak{H_d+K_2, H_d+K_1}+\mathcal{O}(\delta t^3)},
\end{eqnarray}
where
$H_d\not\in\CC,[\CC,\CC]$
ensures that the
second and third terms in
the argument of the exponential 
cannot produce a term proportional to $H_d$.

\section{Examples} 
\label{sec-example}

In this section,
we shall discuss three examples
of systems where singular controls are present.
We will see how the methods and the results in the previous sections
work.
The first two examples are revisits to
previously analyzed, simple two-dimensional systems.
The first example
allows an optimal singular control.
The second allows
singular controls
but they turn out non-optimal
thanks to the GLC condition.
The third example is a three-dimensional system
where singular controls have not been analyzed in detail.
We will see that
the GLC condition excludes
some singular controls
from being optimal
but the others survive.
However, we will observe that consideration of
only a single singular control is enough.

\subsection{Example 1: Landau-Zener model}

First, let us seek the 
time-optimal controls in
the Landau-Zener model,
\begin{eqnarray}
  H(t) = \omega_0\s^z + u(t) \s^x,
\end{eqnarray}
where $\omega_0>0$ is a fixed parameter and
$u(t)$ is a control variable
satisfying $|u(t)|\le\Omega$,
as in Ref.~\cite{Heg2013}.
We have the drift $H_d=\omega_0\s^z$ and
the control subspace
$\CC=\Span\brac{\s^x}$.
The set $\AA$ of available Hamiltonians is typical and lotus leaf type.
From eq.~\eqref{eq-singular-control}, 
a control is singular if
\begin{eqnarray}
  \tr[\s^xF] &= 0.
  \label{eq-lz-sing}
\end{eqnarray}
The time derivatives
of this conditions 
are
\begin{eqnarray}
  \diff{t} \tr[\s^xF]
  &= -i\tr[\s^x[H,F]]
  = -2\omega_0\tr[\s^yF] = 0,\\
  \f{\mathrm{d}^2}{\mathrm{d}t^2}
  \tr[\s^x F]
  &=2i \omega_0\tr[\s^y[H,F] ]
  =4\omega_0u(t)\tr[\s^zF ]= 0.
  \label{eq-lz-sing-opt-2ndderiv}
\end{eqnarray}
If the singular control is time optimal,
we have
from \eqref{eq-singular-control2} that 
\begin{eqnarray}
  \tr\brak{H_dF}
  =
  \omega_0\tr[\s^z F] &= 1. 
  \label{eq-lz-sing-opt}
\end{eqnarray}
Eqs.~\eqref{eq-lz-sing-opt-2ndderiv}
and \eqref{eq-lz-sing-opt}
imply $u(t)=0$.
This is the only singular control that is potentially time optimal.
The GLC condition does not exclude this protocol, which can be
verified through eqs.~\eqref{eq-GLC-mat-tau-concrete}.
Ref.~\cite{Heg2013} actually provides a case,
though it is for the
evolution of quantum states,
that
the ``bang-off-bang'' control
is optimal, where
the ``bang'' control $u(t)=\pm\Omega$ is
regular and the ``off'' control $u(t)=0$ is
singular.

\subsection{Example 2: one-qubit system}

Next, we discuss the one-qubit example 
raised at the end of Sec.~\ref{sec-setup}.
The Hamiltonian reads
\begin{eqnarray}
  H(t) = \omega_0\s^z + u^x(t)\s^x + u^y(t) \s^y,
\end{eqnarray}
where $\omega_0>0$ is a fixed parameter and
$u^x(t)$ and $u^y(t)$ are control variables
satisfying $(u^x)^2+(u^y)^2\le\Omega$.
We have the drift
$H_d=\omega_0\s^z$ and
the control subspace
$\CC=\Span\brac{\s^x,\s^y}$.
The constraint $\AA$ is typical and lotus leaf type.

The singularity condition for $(H(t),F(t))$
is given by $\tr[\CC F]=0$, or
\begin{eqnarray}
  \tr\brak{\s^xF}=\tr\brak{\s^yF}=0.
  \label{eq-ex2-sing}
\end{eqnarray}
We can show that the singular control
has $u^x=u^y=0$ in a way similar to the previous example.
If this singular protocol is time optimal, it is necessary that
\begin{eqnarray}
  \tr\brak{\s^zF}=1. 
  \label{eq-ex2-sing-2}
\end{eqnarray}

Here, we show by the GLC condition
that
the singular controls cannot be optimal.
From
eq.~\eqref{eq-GLC-mat-tau-concrete},
where
$h_1 = \s^x$ and $h_2 = \s^y$,
we have the $(1,2)$-component
of the matrix $Q^{(1)}$,
\begin{eqnarray}
  Q^{(1)}_{12} = -i\tr\brak{\brak{\s^x,\s^y}F} = 2 \tr\brak{\s^z F}.
\end{eqnarray}
The GLC condition implies that
this must vanish.
This
contradicts eq.~\eqref{eq-ex2-sing-2}.
Therefore, the singular control $u^x=u^y=0$
is not time optimal.
Although this statement has already
been shown in Ref.~\cite{Alb.DA2015},
the proof has become easier thanks to the
GLC condition.

\subsection{Example 3: symmetric two-qubit system}
\label{sec-ex3}

Our final example is a quantum control in a three-dimensional Hilbert
space.
We adopt a representation by two spins.
Consider a
Hamiltonian
\begin{eqnarray}
  H_2(t) =
  \omega_0 \s^x_1\s^x_2
  + J(t)\s^z_1\s^z_2
  + \sum_{i=1}^3 \f{b^i(t)}{2}\paren{\s^i_1+\s^i_2},
\end{eqnarray}
where $\omega_0>0$ is a fixed parameter
and
$(J,b^1,b^2,b^3)\in\mathbb{R}^4$ are control variables
in the region  $J^2+(b^1)^2+(b^2)^2+(b^3)^2\le\Omega^2$ 
denoted by $\UU\subset\mathbb{R}^4$.
We assume $\omega_0<\Omega$.

The Hamiltonian is symmetric
under the exchange of the spins.
Thus, the symmetric subspace
$\HH_{\mathrm{sym}}$
of the total Hilbert space $\HH$,
or the space of triplet states,
is invariant under the time evolution by $U(t)$.
We hereafter restrict our attention on $\HH_{\mathrm{sym}}$.
This is
a control problem on $\SU(3)$.
On $\HH_{\mathrm{sym}}$, we can rewrite the Hamiltonian $H_2(t)$ as
\begin{eqnarray}
  H(t)
  = \omega_0
  \tilde\Sigma^x
  + J(t)
  \tilde\Sigma^z
  + \sum_i b^i(t)S^i,
  \label{eq-ex3-H'}
\end{eqnarray}
where $\Sigma^x,\Sigma^z,S^1,S^2$, and $S^3$
are
the restrictions
on
$\HH_{\mathrm{sym}}$
of
$\s^x_1\s^x_2,\s^z_1\s^z_2,(\s^x_1+\s^x_2)/2,(\s^y_1+\s^y_2)/2$, and
$(\s^z_1+\s^z_2)/2$,
respectively,
and
the tilde denotes the traceless part on $\HH_{\mathrm{sym}}$. 
Concrete expressions for them are given in \ref{sec-app4}.

For the Hamiltonian \eqref{eq-ex3-H'}, we have
\begin{eqnarray}
  &H_d
  = \omega_0
  \tilde\Sigma^x,
  \q
  H_c(t)= J(t)
  \tilde\Sigma^z
  + \sum_i b^i(t)S^i,
  \\
  &\CC=
  \Span\normalbrac{
    \tilde\Sigma^z,
    S^1, S^2, S^3
  }
  =
  \Span\normalbrac{
    \lambda_1+\lambda_6,\,
    \lambda_2+\lambda_7,\,
    \lambda_3,\,\lambda_8
  },
\end{eqnarray}
where $\brac{\lam_i}_{i=1}^8$ are the Gell-Mann matrices
(\ref{sec-app4}).
The constraint $\AA$ is planar and lotus leaf type.
Note that the constraint is typical
as a two-qubit problem (on $\HH$) but is not so as
a one-qutrit problem (on $\HH\sb{\mathrm{sym}}$).\footnote{
  If one wants to consider a typical time-optimal control on a qutrit,
  replace the constraint
  $J^2+(b^1)^2+(b^2)^2+(b^3)^2\le\Omega^2$
  with
  $8J^2/3+(b^1)^2+(b^2)^2+(b^3)^2\le\Omega^2$,
  which is equivalent to
  $\tr (H_c)^2\le\Omega^2$,
  in the subsequent discussions.
}

Let us identify the singular controls $(H(t),F(t))$.
We expand $F$ by the Gell-Mann matrices,
$F=\sum_{i=1}^{8}f^i\lambda_i$.
The singularity condition~\eqref{eq-singular-control}
and its derivative~\eqref{eq-singular-control-dt1}
imply that
\begin{eqnarray}
  f^1+f^6 = f^2+f^7 =f^3 = f^8 = f^5 &= 0.
  \label{eq-ex3-singular}
\end{eqnarray}
Under these relations, the algebraic condition \eqref{eq-PMP-const}
[or \eqref{eq-singular-control}] leads to
\begin{eqnarray}
f^4 \neq 0.
  \label{eq-ex3-singular-normal}
\end{eqnarray}

In the following, we perform the GLC test.
We discuss here the case that
$(J,b^1,b^2,b^3)$ is in the interior of $\UU$, 
while we show non-optimality of the singular controls
on the boundary of $\UU$ 
in \ref{sec-app3}.
We calculate the matrix
$Q^{(1)}$
by
eqs.~\eqref{eq-GLC-mat-tau-concrete}
with $h_j = S^j$ for $j=1,2,3$
and $h_4 = \tilde\Sigma^z$.
The only nontrivial
components
are
\begin{eqnarray}
  Q^{(1)}_{41} &= - Q^{(1)}_{14} = i4\sqrt{2}f^2, \q
  Q^{(1)}_{42} &= - Q^{(1)}_{24} = -i4\sqrt{2}f^1.
\end{eqnarray}
The GLC condition $Q^{(1)}=0$  requires  $f^1=f^2=0$.
It follows that $f^i=0$ except $i=4$.
Then,
from the conditions $\mathrm{d}f^i/\mathrm{d}t=-i\tr\brak{\lam_i[H,F]}=0$ for $i\neq4$,
we have
\begin{eqnarray}
  b^i = 0, \quad i=1,2,3.
\end{eqnarray}
Therefore, all variables
other than $f^4$ and $J$ are zero
by the GLC condition for
$m=1$.
Next, we calculate $Q^{(2)}$ to obtain
\begin{eqnarray}
  Q^{(2)} = 4
  {\small\bmat{cccc}
  Jf^4 & 0 & 0 & 0 \\
  0 & (\omega_0-J)f^4 &0 & 0 \\
  0 & 0 & 2f^4\omega_0 & 0\\
  0 & 0 & 0 & 0
  \emat{}}.
  \label{eq-glc-mat-ex-2}
\end{eqnarray}
From the GLC condition,
$Q^{(2)}$ must be positive semidefinite,
which implies
\begin{eqnarray}
  0\le f^4, \quad 0\le J\le \omega_0.
  \label{eq-ex3-f-J}
\end{eqnarray}
This is the condition
that singular optimal controls must satisfy.
As a result,
the singular time-optimal
Hamiltonian
has the form
\begin{eqnarray}
  H(t)= \omega_0
  \tilde\Sigma^x
  +J(t)
  \tilde\Sigma^z,
  \label{eq-ex3-opt}
\end{eqnarray}
where $0\le J(t)\le \omega_0$.

However, a further observation shows that
we can restrict the time-optimal singular controls to
\begin{eqnarray}
  H(t)=H_d
  =
  \omega_0\tilde\Sigma^x
\label{eq-H=Hd}
\end{eqnarray}
only,
because we can replace any optimal singular protocol
\eqref{eq-ex3-opt} with
a regular protocol followed by the 
protocol \eqref{eq-H=Hd} 
without changing the time duration.
Assume that
a time-optimal protocol $H(t)$
in the interval $[t_1,t_2]$ is given by
eq.~\eqref{eq-ex3-opt}.
Then,
the unitary operator in $[t_1,t_2]$ is given by
\begin{eqnarray}
  U_{\mathrm{sing}}(t_2,t_1) =
  e^{-i \omega_0 (t_2-t_1)
    \tilde\Sigma^x
  }
  e^{-i\paren{\int_{t_1}^{t_2} dt\; J(t)}
    \tilde\Sigma^z
  },
  \label{eq-singular-unitary}
\end{eqnarray}
because $\tilde\Sigma^z$ and $\tilde\Sigma^x$ commute.
Since
$J<\Omega$,
there exists 
$t_3\in[t_1,t_2]$
such that
\begin{eqnarray}
  \int_{t_1}^{t_2} dt\; J(t) =
   (t_3-t_1)\Omega.
\end{eqnarray}
We can realize the
unitary operator \eqref{eq-singular-unitary}
by setting $J(t)=\Omega$ on $[t_1,t_3]$
and $J(t)=0$ on $[t_3,t_2]$.
Since the control $J=\Omega$ cannot be singular optimal
{}[as was seen in \eqref{eq-ex3-f-J}],
it is regular.
Therefore, we can deform any
time-optimal singular control 
to a regular control plus the singular control
given by eq.~\eqref{eq-H=Hd}.
Thus, we can restrict ourselves to seek
a sequence consisting of regular solutions of {\QMP} and a
singular control $H(t)=H_d$.
In fact, if the target
unitary operator $U_f$ takes the form
\begin{eqnarray}
  U_f= e^{-i\alpha
    \tilde\Sigma^x
  },
\end{eqnarray}
then the singular control
with $H_c=0$ may be optimal
and the time cost is
$\alpha/\omega_0$.

\section{Conclusion and discussions}
\label{sec-conclusion}

We have discussed the problem of
finding the time-optimal control
on quantum systems. 
The problem is specified by a pair 
$(U_f,\AA)$, 
where $\AA\subset\su(N)$ is the 
set of available Hamiltonians 
representing the theoretical or experimental constraints on the system 
and $U_f\in\SU(N)$ is the target unitary operation. 
Our task is to find 
the Hamiltonian $H(t)\in\AA$
which realizes $U_f$ in the least time. 
Although there has been a formulation by variational principle 
called quantum brachistochrone (QB) 
\cite{Car.Hos.Koi.Oku2006,Car.Hos.Koi.Oku2007}, 
it has a drawback that it is applicable only when 
$\AA$ is expressed by a set of equalities. 
To treat inequality constraints as well, 
we have extended  
QB 
by 
Pontryagin's maximum principle 
which can be viewed as a modern variational calculus. 
The new formulation, which we called 
{\QMP} in this paper, requires the maximization of
a quantity known as 
the Pontryagin Hamiltonian 
with respect to the control
Hamiltonians
at each instant of time.
The solutions to {\QMP} fall into either of the 
two types, 
regular and singular controls.
The singular controls
are those for which
the Pontryagin Hamiltonian
does not depend on the
control Hamiltonian
and therefore we cannot
determine the optimal control
protocol from {\QMP}.
To overcome this issue, we have introduced an 
additional necessary condition for the optimum, 
that is, the generalized Legendre-Clebsch (GLC)
condition. 
We have rewritten the condition as a form suitable for 
quantum time-optimal control. 
This enable 
us to restrict the
form of the optimal singular Hamiltonians.  
This has been demonstrated in the examples.
To summarize, we have constructed a general theory and 
a procedure to find
the time-optimal control in quantum systems 
which can take care of inequality constraints and
singular protocols.

Taking advantage of generality of {\QMP}, 
we have also discussed the relation among 
the drift field, the reduction of inequality constraints, and 
the singular controls. 
To do so, we have classified the general constraint $\AA$ into two 
types, the lollipop type and the lotus leaf type, 
depending on the drift $H_d$ belongs to the control subspace $\CC$ or
not. 
We have also introduced a useful class of constraints, 
called {planar} constraints, 
which 
covers most of commonly encountered situations. 
We have shown the following. 
\begin{enumerate}
\item 
If the constraint $\AA$ is lollipop type, 
there do not exist time-optimal singular controls 
(Sec.~\ref{sec-sing-def}). 
\item 
If the constraint $\AA$ is planar, 
all {regular} optimal controls lie on the boundary of $\AA$ 
(in the control hyperplane $H_d+\CC$), i.e., 
attain an equality among the inequality constraints 
(Sec.~\ref{sec-lollipop} and Sec.~\ref{sec-lotus-leaf}).
\end{enumerate}
From (i) and (ii), 
if $\AA$ is planar and the lollipop type, 
all optimal controls
are regular 
and attain an equality.
On the other hand, 
if $\AA$ is lotus leaf type, 
such reduction depends on the situation 
 (see Sec.\ref{sec-example}). 

Therefore, an efficient 
procedure to 
find the time-optimal control protocol 
for the system with inequality constraints 
is as follows. 
First, check whether the system constraint
is lollipop type or lotus leaf type.
If the constraint is lollipop type,
the optimal control is regular.
Otherwise, the optimal control possibly
consists of regular and singular controls.
The regular optimal controls attain an equality
and one can identify the time-optimal protocol by the QB equation
as in the previous studies~\cite{Car.Hos.Koi.Oku2007}.
Because the singular optimal controls are not identified by 
the maximum principle, 
one must carry out the GLC test. 
To summarize, 
the time-optimal control is 
a solution of the QB equation (regular control) 
for lollipop-type constraints
and
is a certain sequence 
consisting of solutions to QB equation (regular controls)
and singular controls that satisfy the GLC condition
for lotus-leaf-type constraints.

By {\QMP} and the GLC condition, 
we also have 
provided an intuitive understanding that 
the regular control
takes
advantage of
the control Hamiltonian
whereas
the singular control
takes advantage of the
drift Hamiltonian. 
This is best seen 
in the systems with typical constraints: 
the regular controls must have the
maximal norm of the control
Hamiltonian; 
singular controls exist 
only when the 
drift field $H_d$ is 
not in 
$\CC$
or 
$[\CC,\CC]$ 
so that 
the control Hamiltonian 
$H_c\in\CC$ 
cannot be of much help to $H_d$. 
The examples in the last section 
had such singular controls.
Although a similar argument is possible for planar constraints, 
whether such 
an interpretation is possible 
in general
is open.

\section*{Acknowledgments}
H.W. thanks the support from JSPS KAKENHI Grant Number JP19J13010.
T.K.
thanks
the support from
MEXT Quantum Leap Flagship Program (MEXT Q-LEAP) Grant Number
JPMXS0118067285.

\appendix

\section{Construction of tent and its convexity }
\label{sec-app2}

In this appendix,
we shall construct the
tent $\tilde \RR_{(T,U(T))}$ of $\RR$ at $(T,U(T))\in\MM$
mentioned in
Sec.~\ref{sec-app1}
and prove its convexity.

We first show that the needle variations form a space which is
closed under addition and non-negative scalar
multiplication.
These operations are intuitively the corresponding operations
on intervals of the variations.
A simple needle variation
$M(\ta;\d\ta;K)$
is defined in eq.~\eqref{eq-McShane-simple}.
The sum
$M(\ta_1;\d\ta_1;K_1)+
M(\ta_2;\d\ta_2;K_2)$
of needle variations
is defined simply by the composition of these operations if
$\ta_1\ne\ta_2$.
If $\ta_1=\ta_2=:\ta$, the sum is given by arranging them ``side by side,''
namely, the resulting $H'(t)$ is given by
\begin{eqnarray}
  H'(t)=\left\{
  \begin{array}{ll}
    K_1, &  \ta-\d\ta_1-\ta_2<t\le \ta-\d\ta_2, \\
    K_2, &  \ta-\d\ta_2<t\le \ta, \\
    H(t), & \mbox{otherwise}.
  \end{array}
  \right.
\end{eqnarray}
The sum of the sums of simple needle variations, etc., are defined in a
similar manner.
The space of needle variations thus constructed 
is closed under ``addition,''
which
is not commutative.
A non-negative scalar multiplication is defined by
scalar multiplications on all the intervals.
For example, we have
 $\la \paren{
   M(\ta_1;\d \ta_1;K_1)+
   M(\ta_2;\d \ta_2;K_2)
 }
=
M(\ta_1;\la\d \ta_1;K_1)+
M(\ta_2;\la d \ta_2;K_2)
$.
Under these operations,
the space of needle variations is closed.
Including variations of $\d T$ defined in Sec.~\ref{sec-app1},
The space of variations is still closed under addition and non-negative
scalar multiplication.

Second, although the ``addition'' of needle variations of $H(t)$
are non-commutative,
that of the resulting variations of $U(T)$ are commutative.
This can be seen by the fact that the variation of $U(T)$ caused by
$M(\ta;\d \ta_1;K_1)+M(\ta;\d \ta_2;K_2)$ is
\begin{eqnarray}
  \delta U(T) =
  -iU(T,\ta)\brak{\d \ta_1(K_1-H(\ta))+\d \ta_2(K_2-H(\ta))}U(\tau),
\end{eqnarray}
which is commutative.
Thus, the first-order variations $\d\brak{U(T)}$
with respect to the needle variations and the final time
variation
form a convex-linear space.
This convex cone is the tent $\tilde R_{(T,U(T))}$.

\section{Invariance of the GLC condition under
coordinate transformation}
\label{app-coordi-trans}

The matrix
$Q^{(m)}$ depends on
parametrizations of the
same Hamiltonians.
We shall show that
the conditions $Q^{(M)}=0$,
$Q^{(M)}\ge0$, or $Q^{(M)}\le0$
are nevertheless equivalent in any parametrization.

Let
$\bm u=\brac{u^j(t)}$ and $\bm v=\brac{v^j(t)}$
be parametrizations
of the same Hamiltonian $H(t)$.
We have
\begin{eqnarray}
  \pdiff{u^i}&
  =
  \sum_{j=1}^l
  \f{\partial v^j}{\partial u^i}
   \pdiff{v^j},
   \label{eq-app-transform}
\end{eqnarray}
where the Jacobi matrix $\partial v^j/\partial u^i$
is invertible.

First, we will derive
the
relation between
$Q^{(M)}(\bm{u})$ and
$Q^{(M')}(\bm{v})$,
where
we define $M,M'$
are the smallest value
of $m,m'$ for which
$Q^{(M)}(\bm{u})$,
$Q^{(M')}(\bm{v})$ have at least
one nonzero element respectively.
We can assume $M\le M'$
without loss of generality.
Direct application of 
\eqref{eq-app-transform}
to \eqref{eq-GLC-mat} leads to 
\begin{eqnarray}
  Q^{(M)}_{ij} (\bm{u}) &=
  \sum_{k,l=1}^l
  \f{\partial v^k}{\partial u^i}
   \pdiff{v^k}
   \brak{
   \paren{\diff{t}}^M\paren{
   \f{\partial v^l}{\partial u^j}
   \pdiff{v^l}
    \hpmp}} \nonumber\\
&=\sum_{k,l=1}^l
  \f{\partial v^k}{\partial u^i}
  \f{\partial v^l}{\partial u^j}
   \pdiff{v^k}
   \brak{
   \paren{\diff{t}}^M\paren{
   \pdiff{v^l}
    \hpmp}}\nonumber\\
    & \hspace{-3em} 
    +
    \sum_{n=1}^{M}
      \binom{M}{n}
    \sum_{k,l=1}^l
  \f{\partial v^k}{\partial u^i}
   \pdiff{v^k}
   \brak{\paren{\diff{t}}^{n}\paren{\f{\partial v^l}{\partial u^j}}
   \paren{\diff{t}}^{M-n}\paren{
   \pdiff{v^l}
    \hpmp}}.
\end{eqnarray}
However,
the last terms vanish
because
we have
$Q^{(m)}(\bm{v})=0$ for $m<M'$
and
\begin{eqnarray}
\paren{\diff{t}}^{m}
\paren{\pdiff{v^l}\hpmp}
=0
\end{eqnarray}
from eq.~\eqref{eq-singular-control-dt}.
Therefore,
we obtain
\begin{eqnarray}
  Q^{(M)}_{ij} (\bm{u}) &=
  \f{\partial v^k}{\partial u^i}
  Q^{(M)}_{kl} (\bm{v})
  \f{\partial v^l}{\partial u^j}.
\end{eqnarray}
Since the matrix
$\partial v^j/\partial u^i$
is invertible,
we have
$Q^{(M)}(\bm{u})=0\Leftrightarrow
Q^{(M)}(\bm{v})=0$,
$Q^{(M)}(\bm{u})\ge0\Leftrightarrow
Q^{(M)}(\bm{v})\ge0$, and
$Q^{(M)}(\bm{u})\le0\Leftrightarrow
Q^{(M)}(\bm{v})\le0$.

\section{Recurrence relation of the matrix $Q^{(m)}$}
\label{app-Q}

We shall demonstrate the calculation
of the matrix $Q^{(m)}$ in our
formulation
in Sec.~\ref{sec-pmp-singular}.

First, we obtain the
recurrence relation for general
$m>0$ as
\begin{eqnarray}
  Q^{(m)}_{ij}&=
  \pdiff{u^i}\brak{\diff{t}
    \paren{\diff{t}}^{m-1}
    \pdiff{u^j} \hpmp 
  } \nonumber \\
  &=
  \pdiff{u^i}\brak{
    \diff[u^k]{t}
    \pdiff{u^k}
    \paren{\diff{t}}^{m-1}
    \pdiff{u^j} \hpmp 
    + \tr\brak{\diff[F]{t}
      \pdiff{F}
      \paren{\diff{t}}^{m-1}
      \pdiff{u^j} \hpmp}} \nonumber \\
  &=
  \diff{t}
  Q^{(m-1)}_{ij}
  -i\tr\brak{
    \brak{\pdiff[H]{u^i},F}
    R^{(m-1)}_{j} },
  \label{eq-proof-Q-recur}
\end{eqnarray}
where we define
\begin{eqnarray}
  R^{(m)}_{j} :=
  \pdiff{F}
  \paren{\diff{t}}^{m}
  \pdiff{u^j} \hpmp.
\end{eqnarray}
For $m=0$,
we can calculate
\begin{eqnarray}
  R^{(0)}_j
  =\pdiff{F}
  \tr\brak{\pdiff[H]{u^j}F}
  = \pdiff[H]{u^j}, 
\end{eqnarray}
and
can similarly show
\begin{eqnarray}
  R^{(m)}_{j} &=
  \diff[u^k]{t} \pdiff{u^k}R^{(m-1)}_j
  +\pdiff{F}
  \tr\brak{\diff[F]{t}
  R^{(m-1)}_j
  } \nn
  &=
  \diff[]{t} R^{(m-1)}_j
  -i
  \brak{R^{(m-1)}_j,H},
\end{eqnarray}
for general $m>0$.

If $Q^{(k)}=0$ for $k=1,2,\dots,m-1$
for a certain $m>0$,
then
eq.~\eqref{eq-proof-Q-recur} becomes
\begin{eqnarray}
  Q^{(m)}_{ij} = -i\tr\brak{
  \brak{\pdiff[H]{u^i},F}
  R^{(m-1)}_{j}}.
\end{eqnarray}
For $m=1$, we have
\begin{eqnarray}
  Q^{(1)}_{ij}
  &= -i\tr\brak{\brak{\pdiff[H]{u^j},\pdiff[H]{u^i}}F}.
  \label{eq-proof-Q1}
\end{eqnarray}
For $m=2$, we have
\begin{eqnarray}
  Q^{(2)}_{ij}
   &=-i
   \tr\brak{
   \brak{\diff{t}\paren{\f{\partial H}{\partial u^j}},\f{\partial H}{\partial u^i}}
   F}
   -
   \tr\brak{
   \brak{\f{\partial H}{\partial u^i},F}
   \brak{\f{\partial H}{\partial u^j},H}
   }.
\end{eqnarray}
The first term in $Q^{(2)}_{ij}$ above
vanishes when
$\partial H/\partial u^j$ is
time independent.

\section{Operators appearing in Example 3}
\label{sec-app4}
We list some operators appearing in Example 3
(Sec.~\ref{sec-ex3}).

The Gell-Mann matrices are defined as
\begin{eqnarray}
  \lambda_1 &=
  {\small\bmat{ccc}
  0 & 1 & 0 \\
  1 & 0 & 0 \\
  0 & 0 & 0
  \emat{}},
  \quad
  \lambda_2 =
  {\small\bmat{ccc}
  0 & -i & 0 \\
  i & 0 & 0 \\
  0 & 0 & 0
  \emat{}},
  \quad
  \lambda_3 =
  {\small\bmat{ccc}
  1 & 0 & 0 \\
  0 & -1 & 0 \\
  0 & 0 & 0
  \emat{}},\nn
  \lambda_4 &=
  {\small\bmat{ccc}
  0 & 0 & 1 \\
  0 & 0 & 0 \\
  1 & 0 & 0
  \emat{}},
  \quad
  \lambda_5 =
  {\small\bmat{ccc}
  0 & 0 & -i \\
  0 & 0 & 0 \\
  i & 0 & 0
  \emat{}},
  \quad
  \lambda_6 =
  {\small\bmat{ccc}
  0 & 0 & 0 \\
  0 & 0 & 1 \\
  0 & 1 & 0
  \emat{}},\nn
  \lambda_7 &=
  {\small\bmat{ccc}
  0 & 0 & 0 \\
  0 & 0 & -i \\
  0 & i & 0
  \emat{}},
  \quad
  \lambda_8 =
  {\small\f{1}{\sqrt{3}}
  \bmat{ccc}
  1 & 0 & 0 \\
  0 & 1 & 0 \\
  0 & 0 & -2
  \emat{}}.
\end{eqnarray}
They form the orthonormal basis for $\su(3)$.

The operators on
$\HH_{\mathrm{sym}}$
appeared in
Sec.~\ref{sec-ex3}
has the following forms
in the basis
$\brac{
  \ket{\uparrow\uparrow},
  (\ket{\uparrow\downarrow}+
  \ket{\downarrow\uparrow})/2,
  \ket{\downarrow\downarrow}
}$:
\begin{eqnarray}
  \Sigma^x =
  \matx{
  0 & 0 & 1 \\
  0 & 1 & 0 \\
  1 & 0 & 0
  },
  \quad
  \Sigma^z =
  \matx{
  1 & 0 & 0 \\
  0 & -1 & 0 \\
  0 & 0 & 1
  },\q
  S^1 =
  {\small
    \f{1}{\sqrt{2}}
  \bmat{ccc}
  0 & 1 & 0 \\
  1 & 0 & 1 \\
  0 & 1 & 0
  \emat{}},
\nn
  S^2 =
  {\small
  \f{1}{\sqrt{2}}
  \bmat{ccc}
  0 & -i & 0 \\
  i & 0 & -i \\
  0 & i & 0
  \emat{}},
  \;
  S^3 =
  \matx{
  1 & 0 & 0 \\
  0 & 0 & 0 \\
  0 & 0 & -1
  }. 
\end{eqnarray}
The traceless parts of $\Sigma^i$ on
$\HH_{\mathrm{sym}}$
are given by $\tilde\Sigma^i=\Sigma^i-1/3$ while
$S^i$ are traceless themselves.
By using the Gell-Mann matrices,
we have
\begin{eqnarray}
  \tilde\Sigma^x &= \lambda_4 - \f12\lambda_3+\f1{2\sqrt{3}}\lambda_8,
  \quad
  \tilde\Sigma^z= \lambda_3-\f1{\sqrt{3}}\lambda_8,  \nn
  S^x &= \f1{\sqrt{2}} \paren{\lambda_1+\lambda_6}, \quad
  S^y = \f1{\sqrt{2}} \paren{\lambda_2+\lambda_7}, \quad
  S^z = \f1{2} \paren{\lambda_3+\sqrt{3}\lambda_8}.
\end{eqnarray}

\section{The GLC condition for boundary singular controls in
Example 3
}
\label{sec-app3}

In this appendix,
we shall
show that
\textit{boundary} singular controls
in Example 3
(Sec.~\ref{sec-ex3})
with
\begin{eqnarray}
  J^2+(b^1)^2+(b^2)^2+(b^3)^2=\Omega^2
\end{eqnarray}
are not optimal.
We will reduce the number of control variables and apply the GLC condition.

We begin with recalling the conditions that hold in general.
The singularity condition~\eqref{eq-singular-control}
and its derivative~\eqref{eq-singular-control-dt1}
imply
eqs.~\eqref{eq-ex3-singular},
\eqref{eq-ex3-singular-normal}
and the following:
\begin{eqnarray}
  b^1f^2-b^2f^1 &= 0,
  \label{eq-app-pmp-deriv1}\\
  Jf^2 & = 0, \label{eq-app-pmp-deriv2} \\
  f^1(J-\omega_0) & = 0.
  \label{eq-app-pmp-deriv3}
\end{eqnarray}
The second time derivative implies
\begin{eqnarray}
  b^1f^1+b^2f^2+\sqrt{2}b^3f^4&=0.
  \label{eq-app-pmp-deriv4}
\end{eqnarray}
The algebraic condition \eqref{eq-PMP-const}
leads to
$f^4\ne0$ [eq.~\eqref{eq-ex3-singular-normal}].
These hold for both interior and boundary singular controls.

Let us perform the GLC test.
We shall show by contradiction
that singular optimal controls must have $b^i=0 (i=1,2,3)$
and then examine the optimality of such protocols.

First, we assume $b^3\neq0$.
We regard
the Hamiltonian~\eqref{eq-ex3-H'}
as a function of $b^1$, $b^2$ and $J$,
\begin{eqnarray}
  H(t) = \omega_0\tilde{\Sigma}^x
  +J\tilde{\Sigma}^z
  +\sum_{i=1}^2 b^i(t) S^i
  +b^3(b^1,b^2,J;t)S^3,
\end{eqnarray}
where $b^3(b^1,b^2,J;t)=\pm\sqrt{\Omega^2-J^2-(b^1)^2-(b^2)^2}$.
We have
\begin{eqnarray}
  h_i :=
  \pdiff[H]{b^i} &= S^i
  -\f{b^i}{b^3} S^3
  \quad (i=1,2), \nn
  h_3 :=
  \pdiff[H]{J} &=
  \tilde{\Sigma}^z
  -\f{J}{b^3} S^3.
\end{eqnarray}
We obtain the $3\times3$ matrix $Q^{(1)}$
which has nontrivial components
\begin{eqnarray}
  Q^{(1)}_{13} = -Q^{(1)}_{31}
  &= -4\sqrt{2}f^2,
  \q
  Q^{(1)}_{23} = -Q^{(1)}_{32}
  &= 4\sqrt{2}f^1.
\end{eqnarray}
The GLC condition  $Q^{(1)}=0$ leads to
$f^1=f^2=0$.
These together with
eq.~\eqref{eq-app-pmp-deriv4}
and $f^4\neq0$ imply
$b^3=0$.
This contradicts the assumption.
Therefore, the singular optimal controls
must have $b^3=0$.

Next, we assume $b^1\ne0$ under $b^3=0$.
We regard
the Hamiltonian~\eqref{eq-ex3-H'}
as a function of $b^2$, $b^3$ and $J$.
Setting
$h_1:=\partial H/\partial b^2,
h_2:=\partial H/\partial b^3$, and
$h_3:=\partial H/\partial J$,
we obtain $3\times3$ matrix $Q^{(1)}$
with nontrivial components
\begin{eqnarray}
  Q^{(1)}_{13} &=
  -Q^{(1)}_{31} =
  4\sqrt{2}\paren{
  \f{b^2f^2}{b^1}+f^1},
\end{eqnarray}
where $b^1 = \pm\sqrt{\Omega^2-J^2-(b^2)^2}$.
Since GLC condition requires them to vanish,
we have $b^1f^1+b^2f^2= 0$.
Thanks to eq.~\eqref{eq-app-pmp-deriv1} and
\eqref{eq-app-pmp-deriv4},
we obtain $f^1=f^2=0$.
However,
$\mathrm{d}f^2/\mathrm{d}t = -i\tr\brak{\lam_2\brak{H,F}}=0$
gives $b^1=0$,
which contradicts the assumption.
Thus, the singular optimal controls must have $b^1=b^3=0$.

Finally, we assume $b^2\ne0$ under $b^1=b^3=0$.
We obtain $f^1=0$ from eq.~\eqref{eq-app-pmp-deriv1}.
However, the time-derivative
$\mathrm{d}f^1/\mathrm{d}t = -i\tr\brak{\lam_1\brak{H,F}}=0$
gives $b^2=0$, which contradicts the assumption.
The singular optimal control must have $b^i=0 (i=1,2,3)$.

As a result, we have only
the case with $J\neq0$.
We regard
the Hamiltonian~\eqref{eq-ex3-H'}
as a function of $b^i$ $(i=1,2,3)$.
We have simple operators
\begin{eqnarray}
  h_i :=
  \pdiff[H]{b^i} = S^i,
\end{eqnarray}
thanks to the condition $b^1=b^2=b^3=0$.
Since the operators $h_i$ are the same as
those of the interior control,
we have $Q^{(1)}_{ij}=0$ and $Q^{(2)}_{ij}$ as the
first $3\times 3$ matrix of
eq.~\eqref{eq-glc-mat-ex-2}.
Therefore, we obtain
the same conclusion~\eqref{eq-ex3-f-J}.
However,
eq.~\eqref{eq-ex3-f-J}
cannot be satisfied because of $J=\Omega>\omega_0$,
as is explained in the text.

\end{document}